\documentclass[a4paper, 12pt]{article}

\pdfoutput=1
\usepackage{cite}
\usepackage[affil-it]{authblk}
\usepackage[para]{footmisc}
\usepackage{sectsty}
\sectionfont{\large}
\subsectionfont{\normalsize}
\subsubsectionfont{\normalsize}
\paragraphfont{\normalsize}

\usepackage{cmap}
\usepackage{amsmath}
\usepackage{amssymb}
\usepackage{mathtools}
\usepackage{graphicx}
\usepackage{fancyhdr}
\usepackage{indentfirst}
\usepackage{array}
\usepackage{subcaption}
\usepackage{booktabs}
\usepackage[dvipsnames]{xcolor}
\usepackage{hyperref}
\hypersetup{
    colorlinks=true,
    linkcolor={red!50!black},
    citecolor={blue!80!black},
    urlcolor={blue!80!black},
    pdftitle={An implicit level set algorithm for hydraulic fracturing with a stress-layer asymptote},
    pdfauthor={A.V. Valov, E.V. Dontsov, A.N. Baykin, S.V. Golovin}
}

\newcommand{\pd}[2]{  \frac{\partial #1}{\partial #2} }
\newcommand{\opd}[2]{ \frac{\mathrm{d} #1}{\mathrm{d} #2} }
\newcommand{\norm}[1]{\left\lVert#1\right\rVert}
\newcommand{\dint}{\,\mathrm{d}}
\newcommand{\wmin}{w_\mathrm{min}}
\renewcommand{\div}{\mathrm{div}\,}

\newcommand{\bn}{\boldsymbol{n}}           
\newcommand{\bw}{\boldsymbol{w}}           
\newcommand{\bp}{\boldsymbol{p}}           
\newcommand{\bq}{\boldsymbol{q}}           
\newcommand{\bsigma}{\boldsymbol{\sigma}}  
\newcommand{\bQ}{\boldsymbol{Q}}           
\newcommand{\bR}{\boldsymbol{R}}           
\newcommand{\bDeltaL}{\boldsymbol{\Delta \mathcal{L}}}     
\newcommand{\bDeltaSigma}{\boldsymbol{\Delta \sigma}}      
\newcommand{\bwmin}{\boldsymbol{w}_{\mathrm{min}}}         

\newcommand{\bbA}{\mathbb{A}}              
\newcommand{\bbC}{\mathbb{C}}              
\newcommand{\bbI}{\mathbb{I}}              

\newcommand{\tildeW}{\tilde{w}}            
\newcommand{\tildeS}{\tilde{s}}            

\newcommand{\Eprime}{E'}           
\newcommand{\Kprime}{K'}           
\newcommand{\Cprime}{C'}           
\newcommand{\muprime}{\mu'}        

\begin{document}

\title{\Large An implicit level set algorithm for hydraulic fracturing with a stress-layer asymptote}

\author[1, 2]{\normalsize A.V. Valov\thanks{a.valov@g.nsu.ru}}
\author[3]{\normalsize E.V. Dontsov\thanks{egor@resfrac.com}}
\author[1, 2]{\normalsize A.N. Baykin\thanks{alexey.baykin@gmail.com}}
\author[1, 2]{\normalsize S.V. Golovin\thanks{s.golovin@g.nsu.ru}}
\affil[1]{\footnotesize Lavrentyev Institute of Hydrodynamics SB RAS, Novosibirsk, 630090, Russia}
\affil[2]{\footnotesize Novosibirsk State University, Novosibirsk, 630090, Russia}
\affil[3]{\footnotesize ResFrac Corporation, Palo Alto, CA 94301, USA}

\date{}
\maketitle
    
\begin{abstract}
    The capability to simulate a hydraulic fracturing process is an essential tool that can be used to optimize treatment design and increase the efficiency of field operations. In most practical cases, hydraulic fractures propagate in a multi-layered rock formation. As a result, there is a need to incorporate the effect of such heterogeneities in fracturing models to achieve an accurate prediction. To capture the layered structure of rocks, a hydraulic fracture simulator typically requires a fine mesh, which leads to a drastic reduction in computational performance. An alternative is to use more sophisticated models that are capable of providing reasonably accurate predictions even on a relatively coarse mesh. In the case of fracture growth modeling, the pivotal component of the simulation is a fracture front tracking algorithm that accounts for the layered structure of the formation. Consequently, this paper aims to extend the established Implicit Level Set Algorithm (ILSA) to account for the effect of multiple stress layers within the tip asymptote. The enhanced front tracking algorithm involves the stress-corrected asymptote that incorporates the influence of stress layers within the near-tip region. To further increase the validity region of the stress-corrected asymptote, the stress relaxation factor is introduced, and its accuracy is examined. The numerical algorithm is validated against the reference semi-analytical solutions as well as experimental observations. In addition, we investigate the sensitivity of the fracture geometry to mesh size to demonstrate that the front tracking algorithm based on the stress-corrected asymptote retains its accuracy on a coarse mesh.
\end{abstract}

\section{Introduction}
    Hydraulic fractures are tensile cracks that are often created in underground rock formations due to the injection of pressurized fluid. The most common technological application of hydraulic fracturing is production enhancement in low and moderate-permeability reservoirs by the creation of high-permeable pathways~\cite{Economides_Reservoir_Stimulation_1989}. Other engineering applications of hydraulic fracturing include preconditioning the rock mass with the aim of improving caveability and fragmentation for block-caving mining operations~\cite{Jeffrey_hydraulic_fracturing_mining_2000, Katsaga_hydraulic_fracturing_mining_2015}, enhancing the injectivity and capacity of geologic carbon storage reservoirs~\cite{Fu_influence_of_HF_on_carbon_storage_2017, Huerta_hydraulic_fracturing_CO2_2020}, and heat production from geothermal reservoirs~\cite{Kumari_HF_geothermal_2018, Cui_geothermal_vertical_well_2022, Akdas_analytical_multiple_HF_geothermal_2022}. Hydraulic fractures also occur naturally as kilometer-long dikes that are driven by magma from deep underground chambers to the Earth’s surface~\cite{Rubin_magma_filled_cracks_1995, Roper_buoyancy_driven_cracks_2007, Rivalta_dikes_models_review_2015}. Many of the aforementioned applications require at least some level of understanding of fracture behavior. Therefore, it is important to have accurate models to have the ability to predict the propagation of hydraulic fractures.

    The history of mathematical models for hydraulic fracturing has its origin in the middle of the last century. One of the pioneering studies include Khristianovich\--Zheltov\--Geertsma\--De Kler (KGD) model~\cite{Khristianovic_KGD_1955, Geertsma_KGD_1969}, Perkins\--Kern\--Nordgren (PKN) model~\cite{Nordgren_PKN_1972, Perkins_PKN_Fracture_1961}, and the radial or penny-shaped model \cite{Geertsma_Radial_1969, Abe_Radial_1976}. KGD model assumes that the fracture propagates under plane strain elastic conditions. PKN model considers a vertical planar fracture with fixed height propagating horizontally and assumes the state of plane strain in each vertical cross-section. It also assumes that the fracture length is much larger than its height. The radial model is applicable only in a homogeneous formation when the wellbore is along the direction of the minimum principal stress and has mostly mathematical importance. In order to allow fracture height growth, Pseudo-3D (P3D) models have been developed~\cite{Palmer_P3D_1983, Settari_P3D_1986}, which are essentially extensions of the PKN model that allow height growth. The P3D models have been extended to the enhanced P3D model~\cite{Dontsov_EP3D_2015}, the stacked-height model~\cite{Cohen_Stacked_height_P3D_2015}, and the layered P3D models~\cite{Zhang_P3D_layered_rock_2017, Zhang_P3D_layered_elasticity_2018}.

    With an increase in computing power, further developments have shifted towards more accurate, fully planar 3D models (PL3D)~\cite{Barree_PL3D_1983, Clifton_PL3D_1981, Peirce_Detournay_ILSA_2008, Lecampion_Zia_Pyfrac_2019}. Such models assume that the fracture footprint is planar and has an arbitrary shape. PL3D models involve coupling between two-dimensional fluid flow and the fully 3D elasticity equation representing the non-local elastic response. Planar 3D models are able to handle specific types of fractures that pseudo-3D models are unable to model. Planar 3D models have gained wide popularity in engineering applications due to their high accuracy compared to pseudo-3D models and computational efficiency compared to fully 3D models~\cite{Cherny_Lapin_non_planar_fracture_2016, Paul_3D_XFEM_non_planar_2018, Baykin_Planar3D_Biot_2018}.

    Most hydraulic fracturing simulators utilize the propagation condition that is based on Linear Elastic Fracture Mechanics (LEFM)~\cite{Rezaei_Fast_Multipole_2019, Chen_Planar3D_explicit_2020}, while some use the cohesive zone model~\cite{Wang_3D_Fracture_cohesive_zone_2012, Baykin_Planar3D_Biot_2018}. However, several studies have employed the universal tip asymptotic solution to locate the fracture front for the purpose of increasing the overall accuracy of the model~\cite{Lecampion_Peirce_Impact_of_near_tip_2013}. One of the pioneering works in this direction is~\cite{Peirce_Detournay_ILSA_2008}, which introduced a robust and accurate approach to tracking the fracture front using an Implicit Level Set Algorithm (ILSA). Extensions as well as alternative implementations can be found in~\cite{Dontsov_Peirce_ILSA_2017, Lecampion_Zia_Pyfrac_2019, Linkov_Universal_Umbrella_2019}. In particular, the study~\cite{Dontsov_Peirce_ILSA_2017} incorporates the universal tip asymptotic solution~\cite{Dontsov_Peirce_universal_asymptotic_2015} into the numerical scheme as a propagation condition. This universal solution accounts for the simultaneous interplay of the effects of fracture toughness, fluid viscosity, and leak-off. The use of the universal asymptotic solution leads to a more accurate numerical solution on a relatively coarse mesh, as was shown in~\cite{Lecampion_Peirce_Impact_of_near_tip_2013} for the case of a radial fracture.
    
    Despite the tip asymptote accounts for multiple physical processes, its use introduces significant errors when the fracture front crosses a stress layer. As shown in~\cite{Dontsov_Front_Tracking_Comparison_2022}, the geologic stress layers have a dominant impact on the accuracy of fracture geometry, and ILSA combined with the universal asymptote demonstrates a notable reduction in accuracy once the stress layers are considered. Therefore, to obtain more accurate results, the effect of layers should be incorporated into the tip asymptotic solution. The paper~\cite{Dontsov_Homogenization_2017} considers such the propagation condition for a plane-strain fracture, where the tip solution that accounts for layers is calculated numerically using a non-singular integral formulation. Recently, an alternative to ILSA for front tracking was proposed in~\cite{Dontsov_MuLTipEl_2022, Dontsov_MuLTipEl_SPE_2022}. The Multi Layer Tip Element (MuLTipEl)~\cite{Dontsov_MuLTipEl_2022, Dontsov_MuLTipEl_SPE_2022} algorithm introduces the concept of the additional fictitious stress to track the fracture front in lieu of the level set methodology and also accounts for the effect of layers within a fracture element.
    To do the latter, MuLTipEl approach uses the concept of an apparent toughness in combination with the universal tip asymptotic solution. The effect of thin layers is incorporated by calculating the effective values for stress and toughness within the tip element. 
    
    As is evident from the previous paragraph, two options have been employed to account for the effect of stress layers in the propagation condition. The first option is to numerically solve for the tip asymptote, while the second method relies on the concept of apparent toughness. The paper~\cite{Valov_stress_corrected_asymptote_2023} proposes 
    an alternative third method, in which the ordinary differential equation (ODE) approximation of the original non-singular integral equation is considered. This option is more accurate compared to the apparent toughness approach, and yet much more computationally efficient compared with the numerical solution. Therefore, the goal of this paper is to incorporate this ODE-based stress-corrected asymptotic solution into ILSA algorithm and to evaluate the resultant improvement for the cases of multiple stress layers in the rock formation.

    The paper is organized as follows. Section~\ref{sec:mathematical_formulation} outlines the governing equations for a planar hydraulic fracture under consideration. Section~\ref{sec:numerical_scheme} provides motivation for using the tip asymptotic solution with stress layers, briefly summarizes the approaches to solve the layer crossing problem, and also contains details of the implementation into ILSA algorithm. Then, Section~\ref{sec:stress_relaxation} evaluates the accuracy of the tip asymptotic solution and provides corrections to increase the validity region of the tip asymptotes. Section~\ref{sec:numerical_examples} presents a series of numerical examples validating the developed algorithm against reference solutions and experimental data. Section~\ref{sec:mesh_sensitivity} provides a sensitivity analysis of the hydraulic fracture geometry to mesh size and illustrates the importance of using the stress-corrected tip asymptote. The details of the numerical scheme and tip leak-off calculation are presented in Appendix~\ref{appendix:discretization} and Appendix~\ref{appendix:tip_leakoff}.

\section{Mathematical formulation}\label{sec:mathematical_formulation}
\begin{figure}
        \begin{center}
            \includegraphics[width=0.8\linewidth]{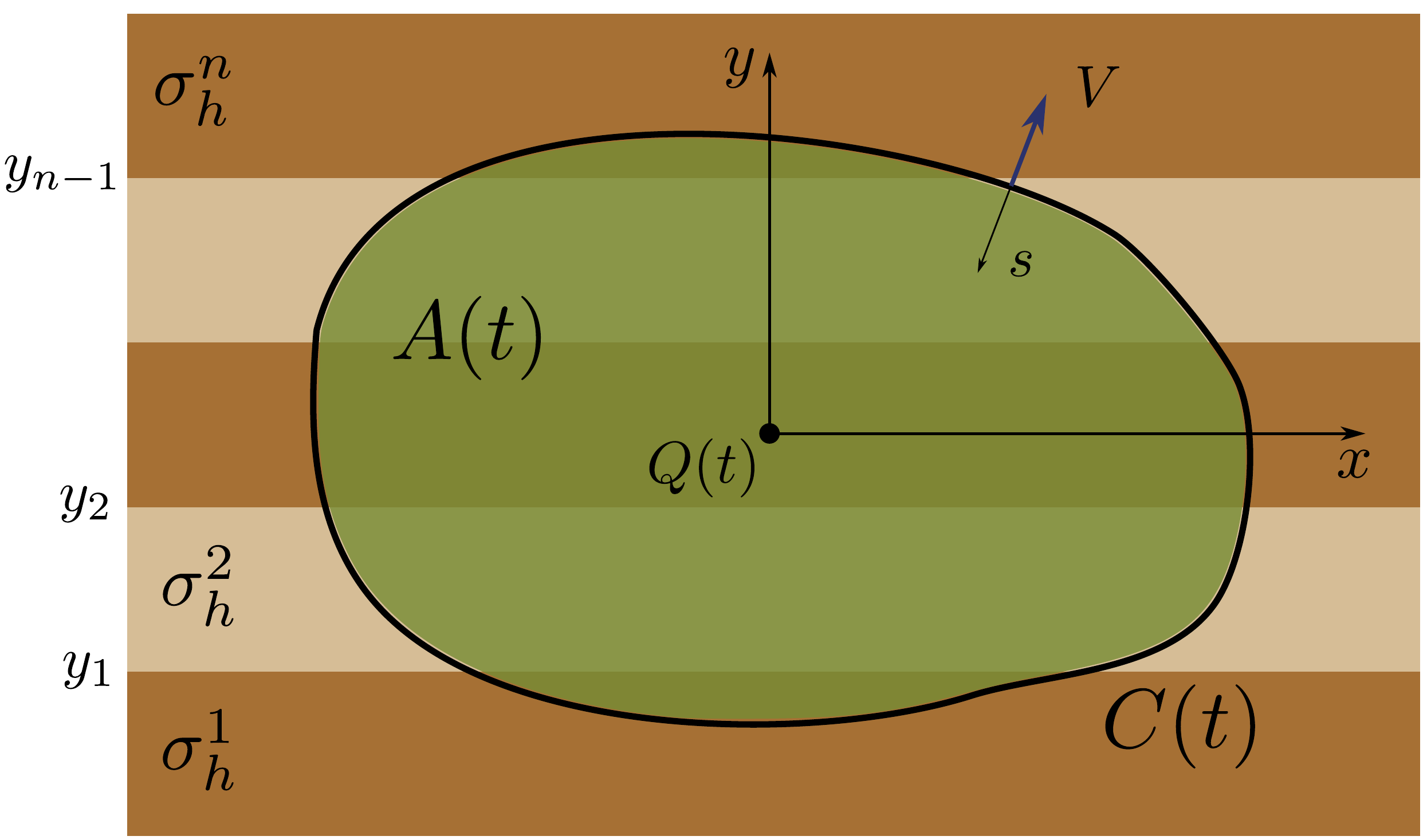}
        \end{center}
        \caption{Scheme of a planar hydraulic fracture with a footprint $A(t)$ bounded by the fracture front $C(t)$ in a layered media. The point source $Q(t)$ is located at the origin. The distance to the fracture front is denoted by $s$ and the normal front velocity is $V$.}
        \label{img:fracture_scheme}
    \end{figure}

    We consider the problem of a planar 3D hydraulic fracture propagating in the $(x, y)$ plane due to the injection of an incompressible viscous fluid with the rate $Q(t)$ through a point source located at the origin, see Figure~\ref{img:fracture_scheme}. The minimum \textit{in situ} compressive stress field $\sigma_h(y)$ is oriented perpendicular to the fracture plane. It is assumed that the fluid front coincides with the fracture front since the fluid lag is negligible for typical depths encountered in field hydraulic fracturing operations~\cite{Garagash_Detournay_Tip_Region_of_a_Fluid-Driven_2000, Lecampion_Implicit_algorithm_fluid_lag_2007}. Fluid leak-off into the surrounding rock is described by Carter's model~\cite{Carter_Leakoff_1957}, fluid rheology is Newtonian, and the effect of hydrostatic pressure is neglected for simplicity. The formation is assumed to be elastically homogeneous, while the stress is allowed to vary from one layer to another. Other parameters, such as toughness and leak-off are assumed to be homogeneous as well for simplicity. We refer to~\cite{Peirce_Detournay_ILSA_2008, Dontsov_Peirce_ILSA_2017, Detournay_Mechanics_of_HF_2016, Lecampion_HF_review_2018} for a more comprehensive description of the underlying assumptions that are typically made in such models. We are looking for the solution for the fracture footprint $A(t)$ and its boundary or front $C(t)$, the fracture width $w$, and the fluid pressure $p$ as a function of the spatial coordinates $(x, y)$ and injection time $t$.

    To simplify mathematical expressions, it is convenient to introduce the following scaled parameters
    \begin{equation*}
        \Eprime = \frac{E}{1-\nu^2}, \qquad \Kprime = 4\left(  \frac{2}{\pi}\right)^{1/2}K_\mathrm{Ic}, \qquad \Cprime = 2C_L, \qquad \muprime = 12\mu,
    \end{equation*}
    where $E$ is the Young's modulus, $\nu$ is the Poisson's ratio, $K_\mathrm{Ic}$ is the fracture toughness, $C_L$ is Carter's leak-off coefficient, and $\mu$ is the fluid viscosity.
    
    \subsection{Elasticity equation}
        The elasticity equation relates the width $w$ of a planar fracture to the normal component of the traction vector $T_n$ applied to the fracture surface. According to the displacement discontinuity method \cite{Crouch_Starfield_BEM_1983, Hills_Kelly_Sol_of_crack_2013}, for the single planar fracture the result can be written in terms of the hypersingular integral equation
        \begin{equation}\label{elasticity_equation}
            T_n(x, y, t) = \sigma_h(y) - \frac{\Eprime}{8\pi} \int_{A(t)} \frac{w(x^\prime, y^\prime, t) \dint x^\prime \dint y^\prime}{\left[(x^\prime - x)^2 + (y^\prime - y)^2\right]^{3/2}}.
        \end{equation}
        Here, $A(t)$ is the fracture footprint enclosed by the fracture front $C(t)$, and $\sigma_h$ is the \textit{in situ} compressive stress field that for the layered formation under consideration varies only with respect to the vertical coordinate $y$. By considering $n$ stress layers in the formation, we can have the following piece-wise constant representation
        \begin{equation*}
            \sigma_h(y) = \sigma_h^1 (1 - \mathcal{H}(y - y_1)) + \sum_{i=2}^{n-1} \sigma_h^i \left[\mathcal{H}(y - y_{i-1}) - \mathcal{H}(y - y_i)\right] + \sigma_h^n \mathcal{H}(y - y_{n-1}),
        \end{equation*}
        where $\mathcal{H}(\cdot)$ denotes Heaviside step function, $\sigma_h^i$ is the stress in the $i$-th layer that is located within the interval $(y_{i-1}, y_i)$, see Figure~\ref{img:fracture_scheme}.
        
        One of the possible scenarios of fracture evolution can be its partial or complete closure. The crack can close, for instance, due to shut-in or decreasing flow rate. Thus, it is necessary to account for the additional constraint for the fracture width. To this end, the fracture width $w$ is restricted to some minimum value $\wmin$ as (see e.g.~\cite{Lecampion_Zia_Pyfrac_2019})
        \begin{equation}\label{contact_condition}
            w(x, y, t) \geq \wmin(x, y, t).
        \end{equation}
        If the fracture is opened $w(x, y, t) > \wmin(x, y, t)$, the normal traction on the fracture surface $T_n$ is equal to the fluid pressure $p$ inside the fracture. Otherwise, the normal traction $T_n$ also accounts for the contact force. Even though the contact algorithm works for any distribution of $\wmin$, this study exclusively focuses on the case without proppant and roughness, for which $\wmin = 0$.
        
    \subsection{Fluid flow}
        Assuming the laminar flow of Newtonian fluid, the fluid flux inside the fracture is reduced to Poiseuille's law
        \begin{equation}\label{poiseuille_law}
            \bq = -\frac{w^{3}}{\muprime}\nabla p,
        \end{equation}
        where $p$ is fluid pressure. By utilizing the assumption that fluid is incompressible, we obtain the continuity equation
        \begin{equation}\label{continuity_equation}
            \pd{w}{t} + \div \bq = Q(t)\delta(x,y) - \frac{\Cprime}{\sqrt{t\!-\!t_0(x,y)}},
        \end{equation}
        where fluid leak-off is calculated according to Carter's model~\cite{Carter_Leakoff_1957}, and $t_0(x,y)$ denotes the time instance at which the fracture front was located at the point $(x, y)$. Substituting Poiseuille's law~\eqref{poiseuille_law} into the mass balance equation~\eqref{continuity_equation} provides the Reynolds lubrication equation
        \begin{equation}\label{reynolds_equation}
            \pd{w}{t} - \div\left(\frac{w^3}{\muprime}\nabla p\right) = Q(t)\delta(x,y) -\frac{\Cprime}{\sqrt{t\!-\!t_0(x,y)}}.
        \end{equation}
        This is a standard equation that describes the evolution of fracture width inside a hydraulic fracture.
        
    \subsection{Boundary conditions}
        We consider the following boundary conditions at the fracture front $C(t)$: zero crack aperture $w = 0$, the propagation criterion $K_{I} = K_\mathrm{Ic}$ and zero flux condition $\bq \cdot \bn = 0$, where $\bn$ is an outward normal to the fracture front in the $(x,y)$ plane. According to the linear elastic fracture mechanics solution for the pure mode I crack propagation~\cite{Rice_Math_Fracture_1968, Peirce_Detournay_Boundary_Cond_2014}, the fracture aperture in the vicinity of the fracture front behaves as
        \begin{equation}\label{propagation_criterion_regular}
            w \underset{s \rightarrow 0}{\sim} \frac{\Kprime}{\Eprime} s^{1/2},
        \end{equation}
        where $s$ is the distance to the fracture font $C(t)$. It is assumed that the fracture propagates in the limit of quasi-static equilibrium, so the normal front velocity is positive $V > 0$. If the fracture doesn't propagate ($V = 0$), then the scaled fracture toughness $\Kprime$ in equation~\eqref{propagation_criterion_regular} should be replaced by the actual scaled stress intensity factor $K_{I}^{\prime}$.
        
\section{Numerical scheme}\label{sec:numerical_scheme}
\begin{figure}
        \centering
        \includegraphics[width=0.9\linewidth]{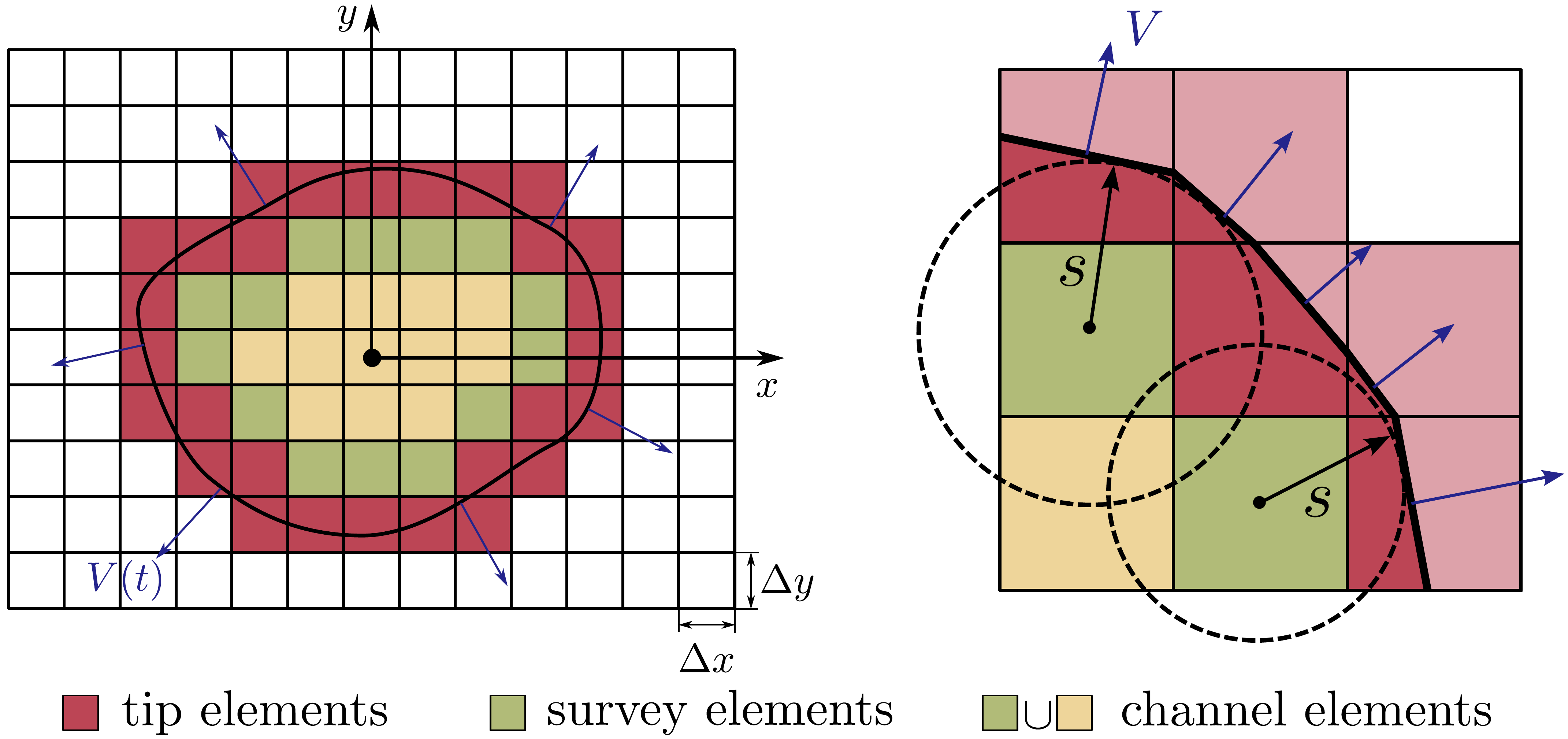}
        \caption{Left panel: Discretization for a planar fracture using a fixed rectangular mesh. Classification of the fracture elements into channel, tip, and survey. The black line shows the fracture front and the arrows schematically indicate the direction of propagation. Right panel: schematic of the fracture front tracking and tip volume calculation, where $s$ is the distance from the center of the survey element to the fracture front. The fluid-filled part of the tip element is highlighted by the dark red color, while the remaining unfilled regions of the tip element are shown by the light red color.}
        \label{img:descrete_mesh}
    \end{figure}

    The hydraulic fracture domain is discretized by a fixed rectangular mesh as shown in Figure~\ref{img:descrete_mesh}. The cell $(i, j)$ is centered at $(x_i, y_j)$ and has dimensions $\Delta x$ and $\Delta y$. All mesh elements are divided into three types: channel, tip, and survey~\cite{Dontsov_Peirce_ILSA_2017}. The channel elements are completely filled with fluid and are contained within the fracture footprint $A(t)$. The tip elements are partially filled with fluid and intersected by the fracture front. The survey elements are a subset of the channel elements and have at least one adjacent tip element.
    
    Backward Euler time stepping is used alongside central differences for spatial discretization. This is a relatively standard approach for such problems~\cite{Peirce_Detournay_ILSA_2008, Dontsov_Peirce_ILSA_2017}. Therefore, details of the numerical scheme and the contact algorithm are given in~\ref{appendix:discretization}. At the same time, the primary focus of this section is to address the issue of fracture front tracking, as is discussed further.
    
    \subsection{The layer crossing problem}
        The ILSA approach invokes the use of the tip asymptotic solution to determine the distance to the fracture front and the fluid volume in the tip elements. One such solution is presented in~\cite{Dontsov_Peirce_universal_asymptotic_2015}, and was used in~\cite{Dontsov_Peirce_ILSA_2017}. This solution simultaneously captures the effects of rock toughness, fluid viscosity, and leak-off in the tip region and can be expressed as
        \begin{equation}\label{universal_asymptotic}
            \frac{s^{2} V \mu'}{E'w_a^{3}} = g_\delta \left(\frac{K' s^{1/2}}{E' w_a}, \frac{2s^{1/2} C'}{w_a V^{1/2}} \right),
        \end{equation}
        where $V$ is the instantaneous fracture front velocity, $w_a(s)$ is the asymptotic solution, $s$ is the distance to the tip, while the definition of the function $g_\delta$ is given in~\cite{Dontsov_Peirce_ILSA_2017}.

        For the given fracture width $w^s$ at the center of a survey element, distance to the fracture front $s$ can be calculated by solving the nonlinear equation~\eqref{universal_asymptotic} with $w_a = w^s$. The fracture front velocity is approximated as $V = (s - s_0) / \Delta t$, where $s_0$ is the distance from the center of the survey element to the fracture front at the previous time step. The distance $s$ is computed for each survey element, and it is used to locate the fracture front as shown on the right panel in Figure~\ref{img:descrete_mesh}. In addition, the fluid in the tip element occupies a polygonal region highlighted by the dark red color. In contrast, the unfilled part of the tip element is highlighted by the light red color. For the purpose of the numerical scheme, the width of the partially filled tip element $w^t$ is calculated from the volume of fluid contained in this element as
        \begin{equation}
            w^t = \frac{1}{\Delta x \Delta y} \int_{PR} w_a(x, y) \dint x \dint y,
        \end{equation}
        where $PR$ is the polygonal region of the tip element occupied by fluid. An effective procedure for integrating the asymptotic solution $w_a$ over the tip element is described in~\cite{Dontsov_Peirce_ILSA_2017}.
        
\begin{figure}
            \centering
            \includegraphics[width=0.65\linewidth]{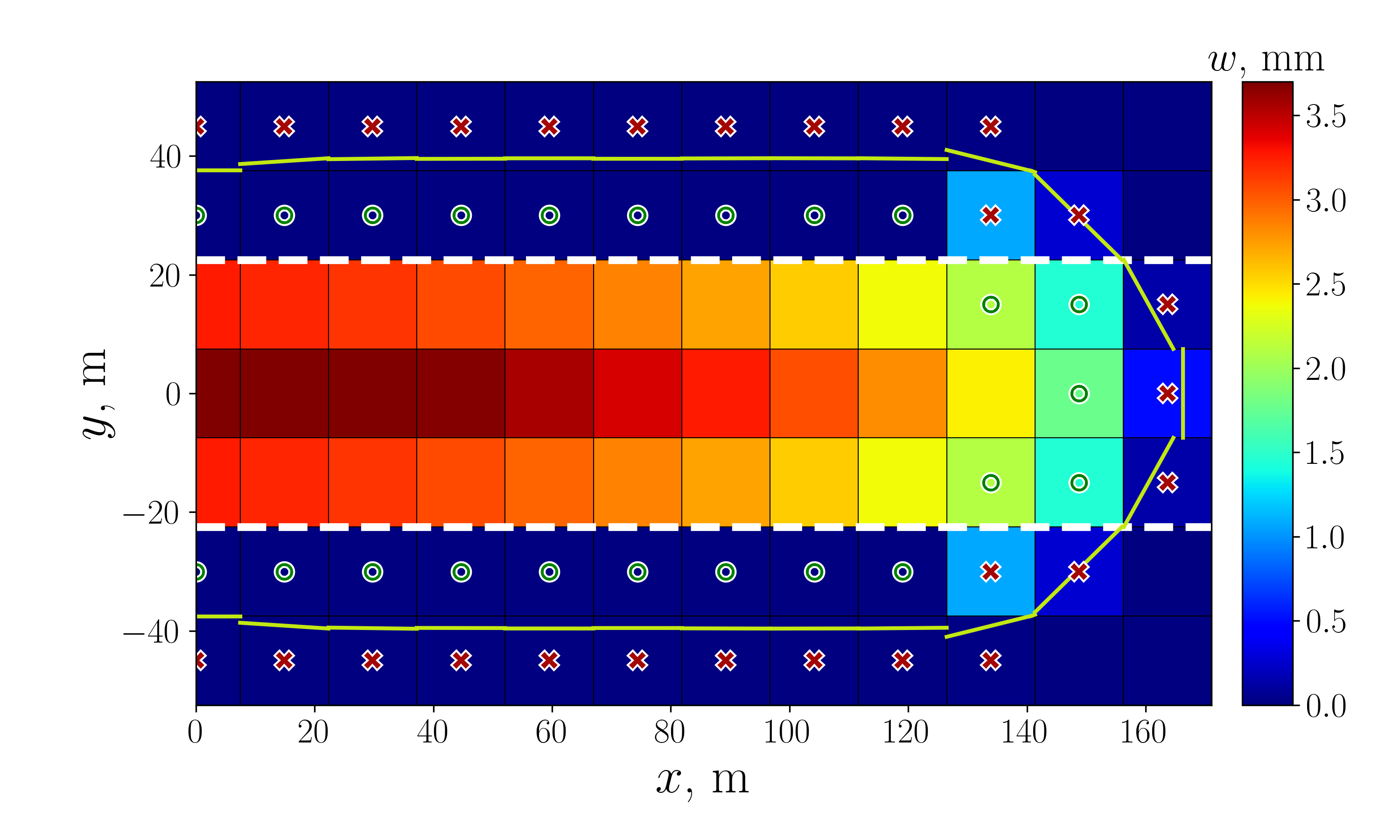}
            \caption{Example numerical calculation of a hydraulic fracture for the case of strong symmetric stress barriers. Dashed white lines show the boundaries of adjacent geological layers, while solid yellow segments indicate the fracture front. Green circular and red crossing markers show the survey and the tip elements, respectively.}
            \label{img:horns_problem}
        \end{figure}

        Despite the aforementioned tip asymptote accounts for multiple physical effects, it may lead to significant errors in situations when the fracture front is crossing a stress layer since the effects associated with layering are not included. The propagation condition is effectively enforced over two numerical elements, from a survey to a tip element. If a geologic layer is located between these two elements, its effect is completely ignored. To illustrate this effect, consider an example with three stress layers where the outer layers have much higher confining stress than the middle layer. This situation corresponds to the propagation of a nearly constant height hydraulic fracture. The results of the calculations are shown in Figure~\ref{img:horns_problem}. The color filling shows the fracture aperture, while the yellow line outlines the fracture front. Dashed white lines indicate the stress layer boundaries. Green circular markers and red crosses signify the survey and the tip elements, respectively. As can be seen from the figure, there is a numerical artifact located near the transition from horizontal to vertical propagation. There are tip elements in the surrounding layers whose width is significantly larger than the width of neighboring elements located in the same layer. This happens because the propagation condition effectively spans over two elements, from a survey to a tip element. And, this propagation condition does not account for the effect of a layer boundary that is located between the two elements.

        To address the problem, the tip asymptotic solution needs to account for the effect of stress layers, or  the so-called layer crossing problem needs to be solved. Such a problem is solved in~\cite{Valov_stress_corrected_asymptote_2023} using several approaches, each having its own advantages and disadvantages. For completeness, this paper provides a brief description of these approaches and primarily focuses on the implementation into the Planar3D ILSA framework.

        \subsection{Tip asymptote from integral equation}
            To address the layer crossing problem, we need to solve the problem of a steadily propagating semi-infinite hydraulic fracture in a stress-layered media. The problem can be mathematically formulated in terms of a non-singular integral equation~\cite{Dontsov_Peirce_universal_asymptotic_2015, Dontsov_Homogenization_2017} as follows
            \begin{equation}\label{integral_equation}
                \begin{aligned}
                    w_a(s) = \frac{\Kprime}{\Eprime} s^{1/2} - &\frac{4}{\pi\Eprime} \int_{0}^{\infty} F(s, x^\prime)\frac{\muprime}{w_a(x^\prime)^2} \left[V + 2\Cprime V^{1/2}\frac{{x^\prime}^{1/2}}{w_a(x^\prime)}\right] \mathrm{d}x^\prime \\ - &\frac{4}{\pi\Eprime} \sum_{j=1}^{n-1} \Delta\sigma_h^j F(s, s_j),
                \end{aligned}
            \end{equation}
            where $s_j$ is the distance from the fracture tip to $j$th stress layer, $\Delta\sigma_h^j = \sigma_h^{j} - \sigma_h^{j+1}$ is $j$th stress jump and the kernel $F(s, x^\prime)$ is given by
            \begin{equation}
                F(s, x^\prime) = (x^\prime - s) \log \left|\frac{s^{1/2} + {x^\prime}^{1/2}}{s^{1/2} - {x^\prime}^{1/2}}\right| - 2 s^{1/2} {x^\prime}^{1/2}.
            \end{equation}
            In order to obtain the solution, the integral in the equation~\eqref{integral_equation} is discretized using Simpson's rule and the obtained system of equations is solved numerically using Newton's method. The infinite upper limit in integral is replaced by a sufficiently large finite number.

            This approach has high accuracy, but at the same time, it is computationally inefficient since it requires solving the system of nonlinear equations on the grid approximating the infinite upper integration limit.

        \subsection{Toughness corrected asymptote}
            An alternative approach attempts to combine the universal asymptotic solution~\eqref{universal_asymptotic} and the integral equation~\eqref{integral_equation} for the case of toughness-dominated propagation by introducing the effective toughness $\Kprime_{\mathrm{eff}}$ as
            \begin{equation}\label{effective_toughness}
                \Kprime_{\mathrm{eff}}(s) = \frac{w_{\mathrm{eff}}(s) \Eprime}{s^{1/2}}, \quad w_{\mathrm{eff}}(s) = \frac{\Kprime}{\Eprime} s^{1/2} - \frac{4}{\pi\Eprime} \sum_{j=1}^{n-1} \Delta\sigma_h^j F(s, s_j),
            \end{equation}
            where $w_{\mathrm{eff}}(s)$ is obtained from the integral equation~\eqref{integral_equation} by excluding the terms responsible for viscous dissipation and fluid leak-off. The effective toughness~\eqref{effective_toughness} is substituted to~\eqref{universal_asymptotic} to yield the asymptotic solution that approximately accounts for stress layers
            \begin{equation}\label{toughness_corrected_solution}
                \frac{s^{2} V \mu'}{E'w_a^{3}} = g_\delta \left(\frac{\Kprime_{\mathrm{eff}}(s) s^{1/2}}{E' w_a}, \frac{2s^{1/2} C'}{w_a V^{1/2}} \right).
            \end{equation}
            The primary advantage of the toughness-corrected approach is the simplicity of implementation and computational efficiency. However, as the influence of viscous dissipation increases, the accuracy of the solution in the presence of stress layers is significantly reduced. In addition, the effective toughness~\eqref{effective_toughness} should be non-negative, which imposes restrictions on the values of $\Delta\sigma_h^j$.

        \subsection{ODE approximation for the asymptote}
            The last solution is based on the ODE approximation of the non-singular integral formulation~\eqref{integral_equation}. By applying the following scaling to the aforementioned integral equation~\eqref{integral_equation}
            \begin{align}\label{scaling_original}
                \begin{split}
                    &\tildeW_a = \frac{\Eprime w_a}{\Kprime s^{1/2}}, \quad 
                    \chi = \frac{2 \Cprime \Eprime}{V^{1/2} \Kprime}, \quad
                    l = \left(\frac{\Kprime^3}{\muprime \Eprime^2 V}\right)^2, \\
                    &\Delta\Sigma_j = \frac{\Delta\sigma_h^j l^{1/2}}{\Kprime}, \quad
                    \tilde{s} = \left(\frac{s}{l}\right)^{1/2}, \quad 
                    \tilde{x}^\prime = \left(\frac{x^\prime}{l}\right)^{1/2},
                \end{split}
            \end{align}
            and differentiating the resulting dimensionless integral formulation, one can substitute the approximation $\tildeW_a \approx \tildeS^\delta$ (as was done in~\cite{Dontsov_Peirce_universal_asymptotic_2015}) to obtain the following initial value problem
            \begin{equation}\label{ode_approx_regular}
                \opd{\widehat{w}_a}{\tildeS} = \frac{\beta_m^3}{3(\widehat{w}_a + G_{\Sigma})^2} + \frac{\chi \beta_{\tilde{m}}^4}{4(\widehat{w}_a + G_{\Sigma})^3}, \qquad \widehat{w}_a(0) = 1,
            \end{equation}
            where
            \begin{equation}\label{ode_stress_term}
                \widehat{w}_a = \tildeW_a - G_{\Sigma}(\tildeS), \quad G_{\Sigma}(\tildeS) = \frac{4}{\pi} \sum_{j=1}^{n} \Delta\Sigma_j \tildeS_j G\left(\frac{\tildeS_j}{\tildeS}\right),
            \end{equation}
            coefficients $\beta_m = 2^{1/3} 3^{5/6}$, $\beta_{\tilde{m}} = 4 / (15 (\sqrt{2} - 1))^{1/4}$, and the function $G(t)$ is defined as
            \begin{equation}
                G(t) = \frac{1 - t^2}{t}\ln\left|\frac{1 + t}{1 - t}\right| + 2.
            \end{equation}
            The initial value problem~\eqref{ode_approx_regular} is solved numerically using explicit Dormand-Prince method~\cite{Dormand_Embedded_RK_methods_1980} with an adaptive step size control~\cite{Macdonald_Adaptive_Step_RK_2001}. The first and the second terms on the right side of the equation~\eqref{ode_approx_regular} allow us to capture the influence of stress layers in the case of non-negligible viscous dissipation and leak-off. Also, there is no need to impose additional restrictions on the magnitude of $\Delta\Sigma_j$ to solve equation~\eqref{ode_approx_regular}. Thus, the ODE approximation is more accurate than the toughness-corrected approach. At the same time, the method is computationally more efficient than the integral equation approach.
            
            Based on the advantages and disadvantages of the approaches described above, we select the ODE approximation as a primary method for fracture front tracking. See also~\cite{Valov_stress_corrected_asymptote_2023} for the detailed comparison of the accuracy and performance of the different approaches.

    \subsection{Locating the fracture front}
        To determine distance $s$ to the fracture front, it is necessary to invert the asymptotic solution $w_a(s, V)$ for the given fracture width $w^s$ of the survey element. Since there is no stress variation in the horizontal direction, the solution without the stress correction is used. At the same time, since the stress varies in the vertical or $y$ direction, then the stress-corrected asymptote needs to be used.
        
        The fracture front velocity $V$ is unknown at the current time step, and it is therefore approximated by the backward Euler scheme as:
        \begin{equation*}
            V = \frac{s - s_0}{\Delta t},
        \end{equation*}
        where $s_0$ is the distance to the fracture front from the previous time step. For the case of vertical propagation, the distance from the $j$th stress layer to the fracture tip $s_j$ is also unknown at the current time step. To overcome this, we introduce an auxiliary parameter $\rho_\mathrm{layer}^j = y_j - y^s$ representing the signed vertical distance from the center of the survey element to the location of the $j$th stress layer, where $y_j$ is the $y$ coordinate of the $j$th stress layer boundary and $y^s$ is the $y$ coordinate of the center of survey element. Therefore, the distance from the fracture tip to the $j$th stress layer is defined as $s_j = s - \rho_\mathrm{layer}^j$. 

\begin{figure}
            \begin{subfigure}[b]{0.48\textwidth}
                \centering
                \includegraphics[width=1.0\linewidth]{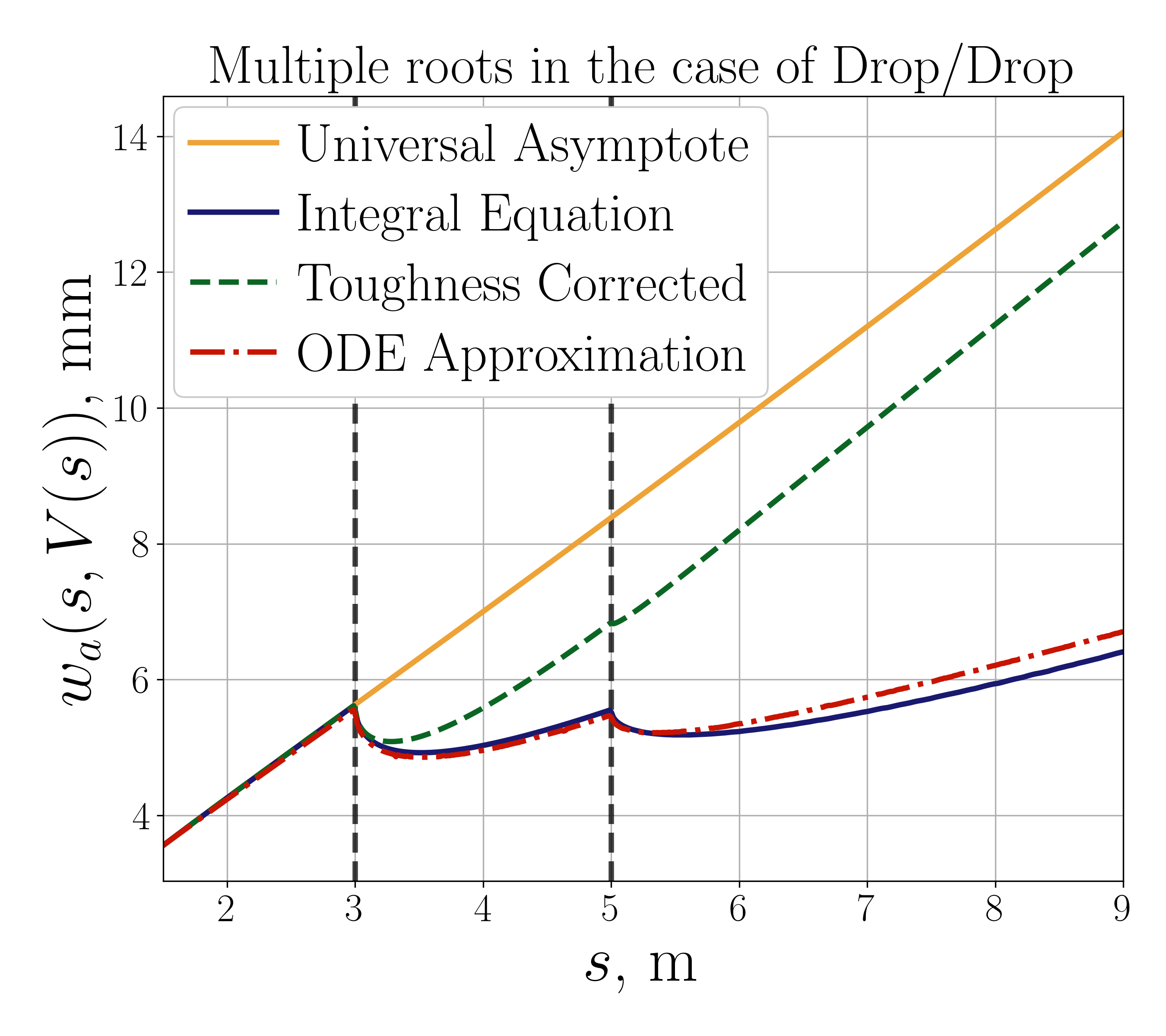}
                \caption{Several roots in the case of two layers}
                \label{img:locating_front:several_roots}
            \end{subfigure}
            \begin{subfigure}[b]{0.48\textwidth}
                \centering
                \includegraphics[width=0.9\linewidth]{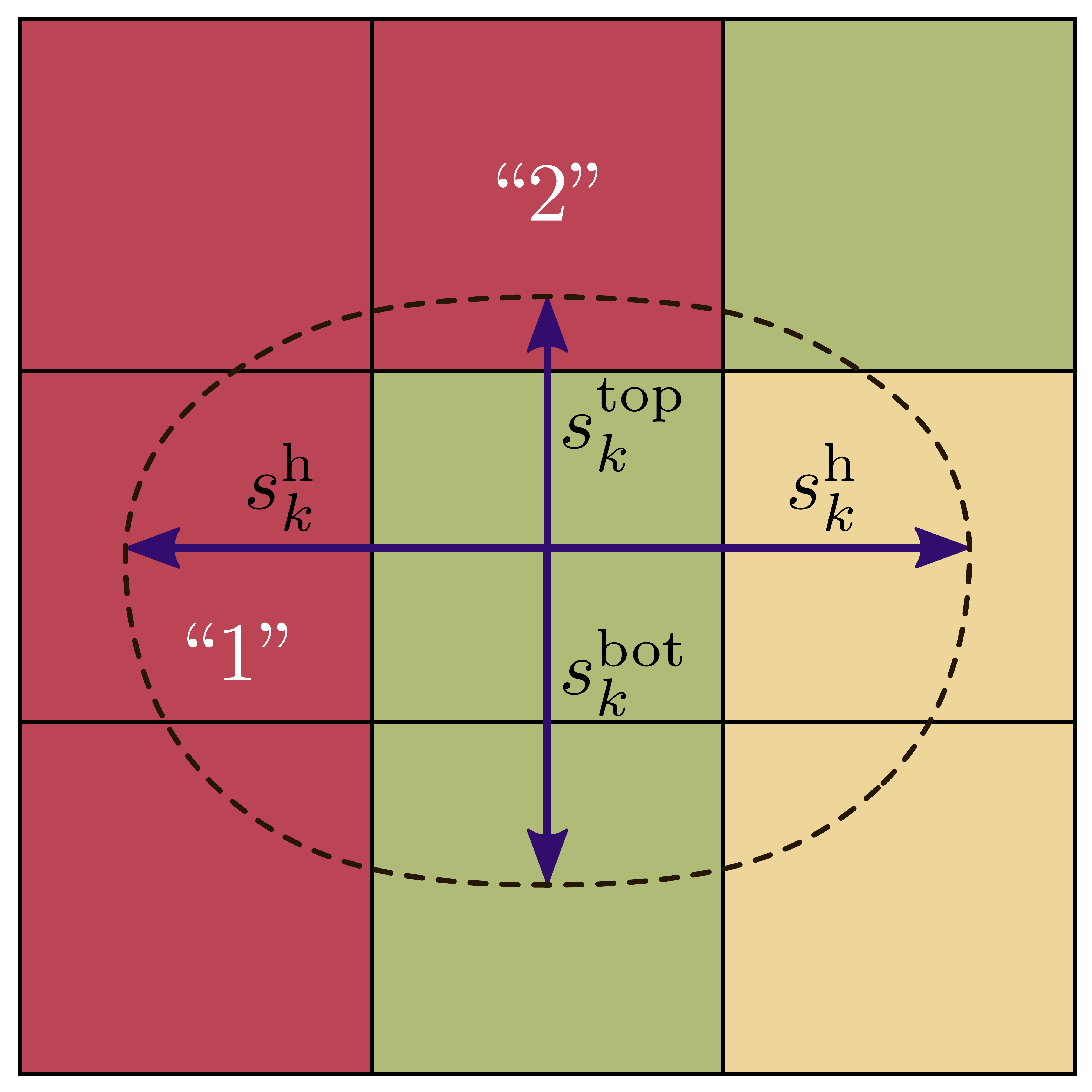}
                \caption{Initial data for the eikonal equation}
                \label{img:locating_front:eikonal}
            \end{subfigure}
            \caption{Panel $(a)$: Comparison of the asymptotic solutions for the case of two stress drop layers located ahead of a survey element. The value of the stress drop is $\Delta\sigma_h^1 = -0.3$ MPa and $\Delta\sigma_h^2 = -0.2$ MPa. Panel $(b)$: Distances in the top, bottom, and horizontal directions from the center of the survey element to the fracture front that are used as an initial condition for the eikonal equation.}
            \label{img:locating_front}
        \end{figure}

        Figure~\ref{img:locating_front:several_roots} shows the comparison of various asymptotic solutions $w_a(s, V)$ for the case of two stress drops located at  $\rho_\mathrm{layer}^1 = 3$~m and $\rho_\mathrm{layer}^2 = 5$~m from the center of the survey element. It is important to point out that the presence of at least one stress drop can make the asymptotic solution non-monotonic. As a result, the equation $w_a(s, V(s)) = w^s$ can have several roots $s_i$ corresponding to different distances to the fracture front for the given width $w^s$ of the survey element. The smallest root is chosen in the algorithm because other roots correspond to the situation of fracture crossing a layer. But if there is a solution before the layer, it means that the fracture is unable to cross the layer and therefore such roots are not physical. In the case of stress jumps or the absence of layers, the asymptotic solution $w_a(s, V(s))$ is monotonic and there is only one root.

        According to ILSA approach~\cite{Dontsov_Peirce_ILSA_2017}, the location of the fracture front is defined by the level set $\mathcal{T} = 0$ of the signed distance function $\mathcal{T}(x, y)$, which is the solution of the eikonal equation~\cite{Dontsov_Peirce_ILSA_2017}. To account for the horizontal stress layers in the plane, we distinguish the top, bottom, and horizontal directions of the fracture propagation. Equation $w_a(s, V(s)) = w^s$ is solved three times for each survey element. First, we need to find a solution assuming that the fracture tip propagates upward to yield distance $s_k^\mathrm{top}$. Then the equation is solved for the distance $s_k^\mathrm{bot}$, assuming that the fracture front propagates downward. And finally, we have to find the distance $s_k^\mathrm{h}$ to the fracture front in the horizontal direction, i.e. without layers. The obtained distances to the fracture front are used as an initial condition to solve the eikonal equation for the signed distance function $\mathcal{T}$ as
        \begin{equation}\label{eikonal_init_data}
            \mathcal{T}(x_k, y_k) = \left\{-s_k^\mathrm{top}, -s_k^\mathrm{bot}, -s_k^\mathrm{h}\right\},
        \end{equation}
        where $(x_k, y_k)$ is the center of the $k$th survey element. For the remaining elements, the signed distance $\mathcal{T}$ takes only one value. To compute the values of $\mathcal{T}$ in the remaining grid cells, the eikonal equation is solved using the fast marching method~\cite{Sethian_FMM_Monotonical_Front_1996, Barentzen_FMM_Implementation_For_3D_2001} and the following finite difference scheme~\cite{Sethian_FMM_Image_Processing_1996}
        \begin{multline}\label{eikonal_upwind_scheme}
            \max{\left(\frac{\mathcal{T}_{i,j} - \mathcal{T}_{i-1,j}}{\Delta x}, \frac{\mathcal{T}_{i,j} - \mathcal{T}_{i+1,j}}{\Delta x}, 0\right)}^2 + \\ \max{\left(\frac{\mathcal{T}_{i,j} - \mathcal{T}_{i,j-1}}{\Delta y}, \frac{\mathcal{T}_{i,j} - \mathcal{T}_{i,j+1}}{\Delta y}, 0\right)}^2 = 1.
        \end{multline}
        In order to account for multiple initial values in each survey element, we modify the finite differences in~\eqref{eikonal_upwind_scheme} as follows:
        \begin{itemize}
            \item If element $(i, j - 1)$ is the $k$th survey, then $\mathcal{T}_{i,j-1} = -s_k^\mathrm{top}$;
            \item If element $(i, j + 1)$ is the $k$th survey, then $\mathcal{T}_{i,j+1} = -s_k^\mathrm{bot}$;
            \item If element $(i \pm 1, j)$ is the $k$th survey, then $\mathcal{T}_{i \pm 1,j} = -s_k^\mathrm{h}$.
        \end{itemize}
        To illustrate the procedure of the finite difference computation, let us consider Figure~\ref{img:locating_front:eikonal}. For the cell ``1'' the horizontal difference takes the form
        \begin{equation*}
            \max{\left(\frac{\mathcal{T}_{i,j} - \mathcal{T}_{i-1,j}}{\Delta x}, \frac{\mathcal{T}_{i,j} - \mathcal{T}_{i+1,j}}{\Delta x}, 0\right)} = \frac{\mathcal{T}_{i,j} + s_k^{\mathrm{h}}}{\Delta x},
        \end{equation*}
        while for the cell ``2'' the vertical difference is calculated using $s_k^\mathrm{top}$ value from the neighboring survey element
        \begin{equation*}
            \max{\left(\frac{\mathcal{T}_{i,j} - \mathcal{T}_{i,j-1}}{\Delta y}, \frac{\mathcal{T}_{i,j} - \mathcal{T}_{i,j+1}}{\Delta y}, 0\right)} = \frac{\mathcal{T}_{i,j} + s_k^{\mathrm{top}}}{\Delta y}.
        \end{equation*}
        A finite difference accounting for the distance $s_k^\mathrm{bot}$ is constructed similarly.
    
    \subsection{Tip volume calculation}
        \begin{figure}
            \begin{center}
                \includegraphics[width=0.7\linewidth]{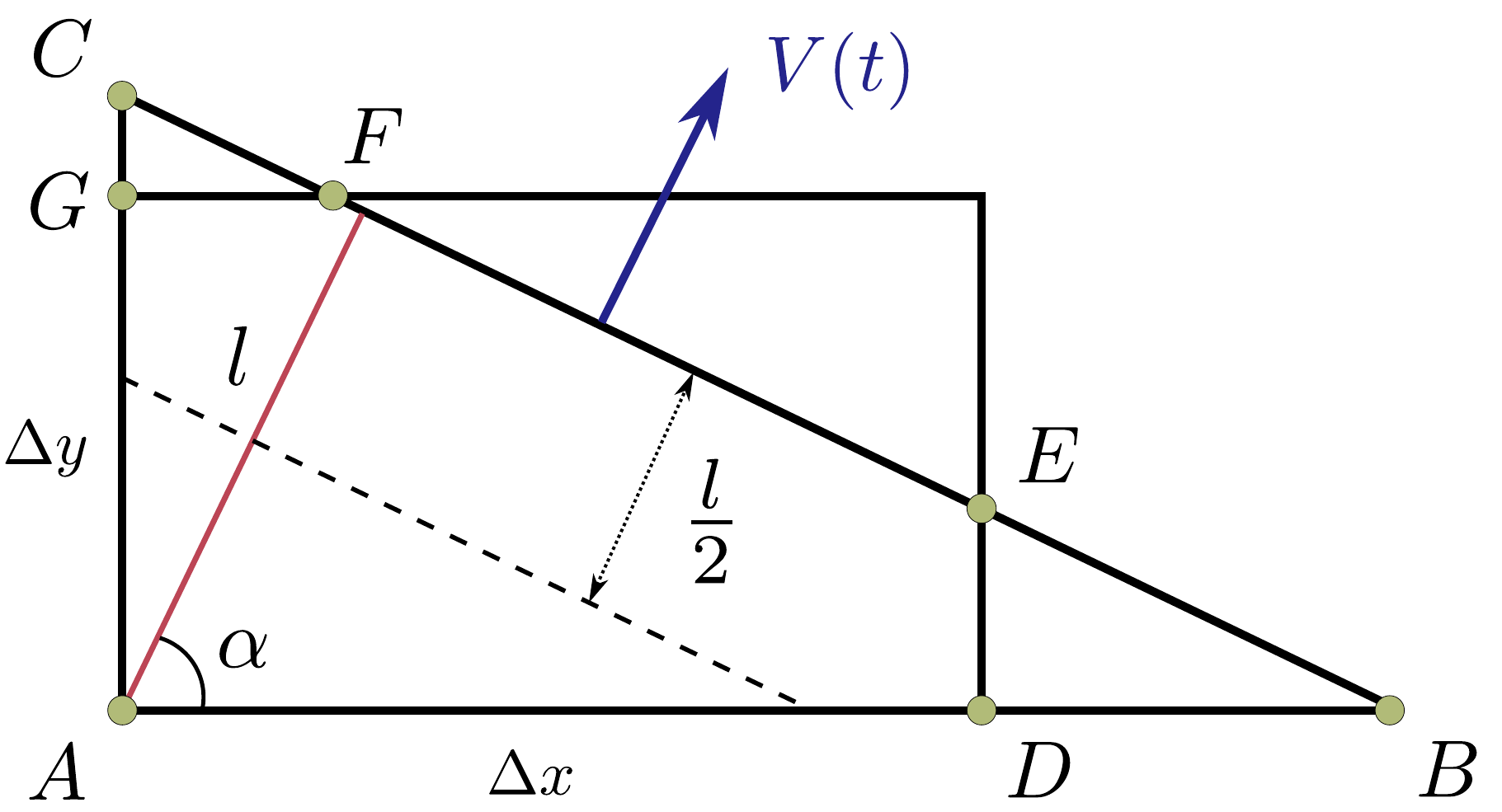}
            \end{center}
            \caption{Scheme to integrate the asymptotic solution over the partially filled tip element.}
            \label{img:tip_element_scheme}
        \end{figure}

        As was mentioned above, the tip width $w^t$ is computed from the volume of fluid occupying the tip element by integrating the asymptotic solution $w_a$ over the partially filled tip element. As can be seen from Figure~\ref{img:tip_element_scheme}, the fluid volume occupying the polygonal region $ADEFG$ can be calculated similarly as in~\cite{Dontsov_Peirce_ILSA_2017}, namely
        \begin{equation}
            V_{ADEFG} = V_{ABC} - \mathcal{H}(\ell - \Delta y \sin\alpha)V_{GFC} - \mathcal{H}(\ell - \Delta x \cos\alpha)V_{DBE},
        \end{equation}
        where $\mathcal{H}(\cdot)$ is the Heaviside step function, $\ell$ is the distance from the farthest interior corner to the front, and $\alpha$ is the front orientation. Note that the triangle $GFC$ disappears when $\ell < \Delta y \sin{\alpha}$, while the triangle $DBE$ disappears when $\ell < \Delta x \cos{\alpha}$. The volume of fluid occupying a triangular region can be computed as
        \begin{equation}
            V_\triangle(s) = \frac{1}{\sin{\alpha}\cos{\alpha}} \int_{0}^{s} s' w_a(s - s') \dint s' = \frac{s M_0(s) - M_1(s)}{\sin{\alpha}\cos{\alpha}},
        \end{equation}
        where $M_0(s)$ and $M_1(s)$ are zeroth and first moments of the asymptotic solution and are defined as
        \begin{equation}
            M_0(s) = \int_{0}^{s} w_a(s') \dint s', \qquad M_1(s) = \int_{0}^{s} s' w_a(s') \dint s'.
        \end{equation}
        Therefore, to calculate the tip volume, it is sufficient to calculate the two moments of the asymptotic solution $w_a$.
        
        Since the stress layers are arranged vertically, they should not affect the fracture tip propagation in the horizontal direction. Thus, in the case of horizontal front orientation or $\alpha = 0$, the fracture tip width is determined by the universal asymptotic solution that we denote as $w_a^u(s)$ in this section. On the other hand, when the fracture propagates in the vertical direction or $\alpha = \pi / 2$, then the influence of the stress layers needs to be included, and the fracture tip width is determined by the asymptotic solution accounting for stress layers that we denote as $w_a^\sigma(s)$ in this section. To account for the arbitrary orientation of the fracture front, an interpolation scheme is implemented as follows
        \begin{equation}\label{interpolated_tip_width}
            w_a(s) = w_a^u(s) \cos^2\alpha + w_a^\sigma(s) \sin^2\alpha.
        \end{equation}
        The zeroth and first moments of the interpolated asymptotic solution~\eqref{interpolated_tip_width} can be expressed as
        \begin{equation}
            \begin{split}
                M_0(s) = \int_{0}^{s} w_a(s') \dint s' = M_0^u(s) \cos^2\alpha + M_0^\sigma(s) \sin^2\alpha, \\
                M_1(s) = \int_{0}^{s} s' w_a(s') \dint s' = M_1^u(s) \cos^2\alpha + M_1^\sigma(s) \sin^2\alpha, \\
            \end{split}
        \end{equation}
        where $M_0^u$ and $M_1^u(s)$ are zeroth and first moments of the universal asymptotic solution~\cite{Dontsov_Peirce_ILSA_2017}, while $M_0^\sigma$ and $M_1^\sigma$ are zeroth and first moments of the asymptotic solution accounting for stress layers. The computation of the moments in the case of the universal asymptotic solution is given in~\cite{Dontsov_Peirce_ILSA_2017}. The moments $M_0^\sigma$ and $M_1^\sigma$ are calculated by numerical integration of the solution with stress layers.

\section{Stress relaxation for the asymptote}\label{sec:stress_relaxation}
    As shown in~\cite{Valov_stress_corrected_asymptote_2023}, the tip asymptote with stress layers (or stress-corrected asymptote) has a relatively small validity region, which may lead to significant errors in quantifying the fracture front location, especially for the case of a coarse mesh. Therefore, the purpose of this section is to evaluate the accuracy of the tip asymptote in the case of a planar fracture. In addition, we consider suitable corrections to increase the validity region of the tip asymptotic solution. 
    
    In particular, we focus on the case when the reservoir layer is surrounded by two identical layers with higher confining stresses (i.e. symmetric stress barriers).  The problem parameters that are used for calculations are the following: Young's modulus $E = 10$ GPa, Poisson's ratio $\nu = 0.3$, fracture toughness $K_\mathrm{Ic} = 1$ MPa$\cdot$m$^{1/2}$, leak-off coefficient $C_L = 0$, fluid viscosity $\mu = 0.1$ Pa$\cdot$s, injection rate $Q = 0.2$ m$^3$/s, height of the central layer $H = 15$ m, and compressive stresses $\sigma_h^1 = \sigma_h^3 = 28$ MPa, and $\sigma_h^2 = 25$ MPa. The computations are performed for the different number of elements in the central reservoir layer $n_c \in \{3, 5, 9, 15\}$, and the mesh is such that $\Delta x = \Delta y = H / n_c$.

    \begin{figure}
        \centering
        \includegraphics[width=1.0\linewidth]{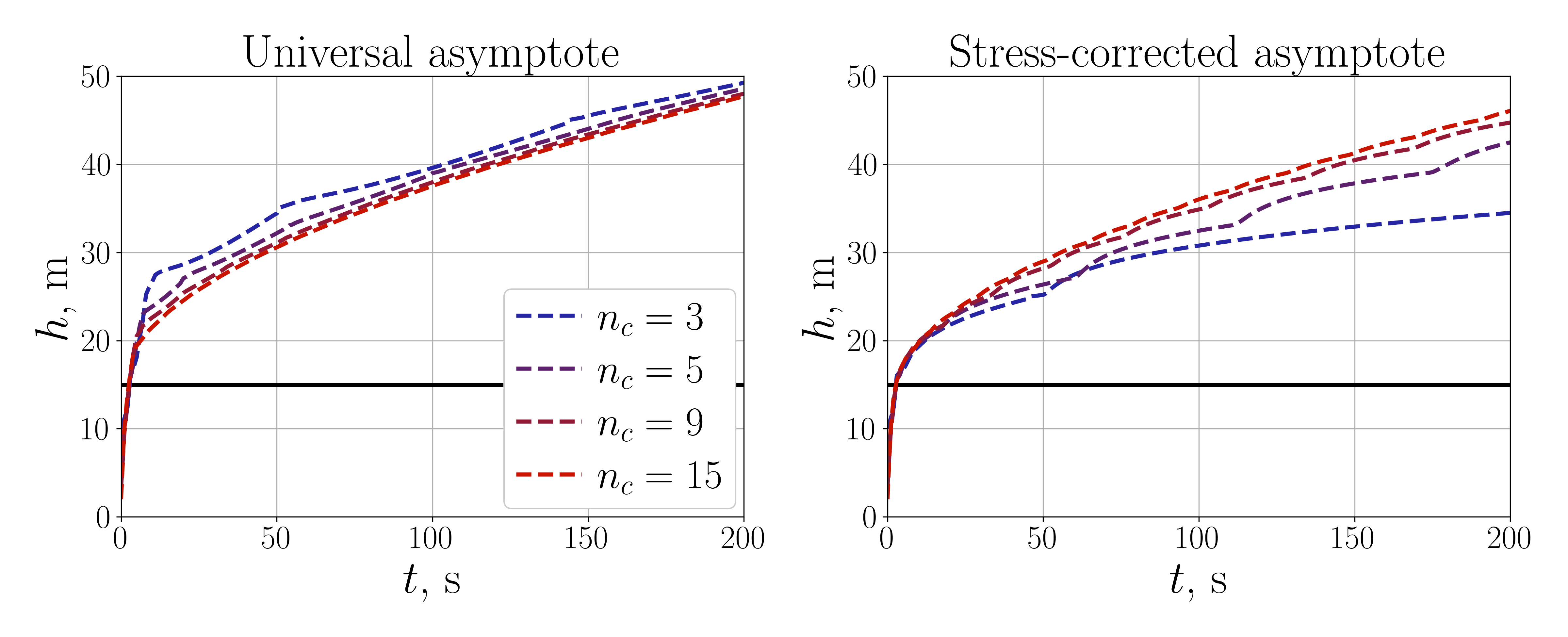}
        \caption{Fracture height versus time for the example case of symmetric stress barriers computed for the different meshes using the universal asymptotic solution (left) and the stress-corrected asymptote (right). The horizontal black solid line indicates the location of the stress layer.}
        \label{img:stress_calibration_height}
    \end{figure}

    Figure~\ref{img:stress_calibration_height} shows the fracture height computed for the different meshes for the universal asymptotic solution~\eqref{universal_asymptotic} and the stress-corrected asymptote calculated via ODE approximation~\eqref{ode_approx_regular}. The horizontal black solid line indicates the location of the high stress layer. As can be seen from the left panel of Figure~\ref{img:stress_calibration_height}, the accuracy of the universal asymptote is compromised near the layer boundary, especially for coarse meshes. However, as the fracture height increases, the mesh size dependence reduces. In contrast, the accuracy of the stress-corrected asymptote is great near the layer intersection, but it is reduced further away from the layer boundary. This example clearly shows that the stress-corrected ODE approximation requires a modification to mitigate the inaccuracy associated with the fracture front being away from the layer boundary.

    The paper~\cite{Valov_stress_corrected_asymptote_2023} proposed a stress relaxation factor $\lambda$ that allows us to significantly extend the validity region of the stress-corrected asymptote. The stress relaxation factor is incorporated into the solution by modifying the stress terms in~\eqref{ode_stress_term} as follows
    \begin{equation}\label{ode_stress_term_corrected}
        G_{\Sigma}(\tildeS) = \frac{4}{\pi} \sum_{j=1}^{n} \lambda(\tildeS_j) \Delta\Sigma_j \tildeS_j G\left(\frac{\tildeS_j}{\tildeS}\right),
    \end{equation}
    where the stress relaxation factor $\lambda$ has values from 0 to 1. Thus, $\lambda$ effectively reduces the magnitude of the stress barrier. For a finite plane-strain fracture with symmetric stress barriers, the optimal relaxation factor has the following form (see~\cite{Valov_stress_corrected_asymptote_2023})
    \begin{equation}\label{lambda_opt_plane_strain}
        \lambda_\mathrm{opt}(\xi) = \frac{0.7}{0.7 + \xi / H},
    \end{equation}
    where $\xi$ is the distance from the fracture tip to the stress layer and $H$ is the height of the central layer. Unfortunately, such a relaxation factor depends on the height of the central layer, which is not well-defined for an arbitrary configuration of the reservoir layers. Therefore, we aim to determine an alternative expression for the stress relaxation factor that does not depend on the geometry of the reservoir layers. 
    
    Results shown in Figure~\ref{img:stress_calibration_height} demonstrate that the stress-corrected asymptote leads to more accurate results near the layer boundary, while the universal asymptote excels far away from the layer boundary. In view of this observation, we are looking for the stress relaxation factor in the following form
    \begin{equation}
        \lambda_k(\xi) = 
        \begin{cases}
            1 - \xi / (k \Delta y), \quad &\xi < k \Delta y, \\
            0, \quad &\xi \geq k \Delta y,
        \end{cases}
    \end{equation}
    where $k$ represents the relaxation distance measured in the number of elements.
    
    To determine the optimal value of the relaxation distance $k$, we consider the case of symmetric stress barriers for the parameters specified at the beginning of this section. Then, we independently vary fluid viscosity $\mu \in [0.01, 0.5]$ Pa$\cdot$s, fracture toughness $K_\mathrm{Ic} \in [0.001, 5]$ MPa$\cdot$m$^{1/2}$, and magnitude of the stress barrier $\Delta\sigma \in [1.7, 5]$ MPa, where $\Delta\sigma = \sigma_h^3 - \sigma_h^2$. To compare the performance of the relaxation factors with different values of relaxation distance $k$, we take the error measure as the average difference between the solution for the fine mesh $n_c = 15$ and the solutions for the coarser meshes as follows
    \begin{equation}\label{eps2}
        \varepsilon_2 = \max_{n_c} \frac{\norm{h_{n_c}(t) - h_{15}(t)}}{\norm{h_{15}(t)}}, \qquad \norm{f(t)}_2^2 = \int_{0}^{t_\mathrm{max}} f^2(t) \dint t,
    \end{equation}
    where the subscript $n_c$ denotes the number of elements in the central layer and $t_\mathrm{max} = 200$ seconds is the calculation time.

    \begin{figure}
        \centering
        \includegraphics[width=1.0\linewidth]{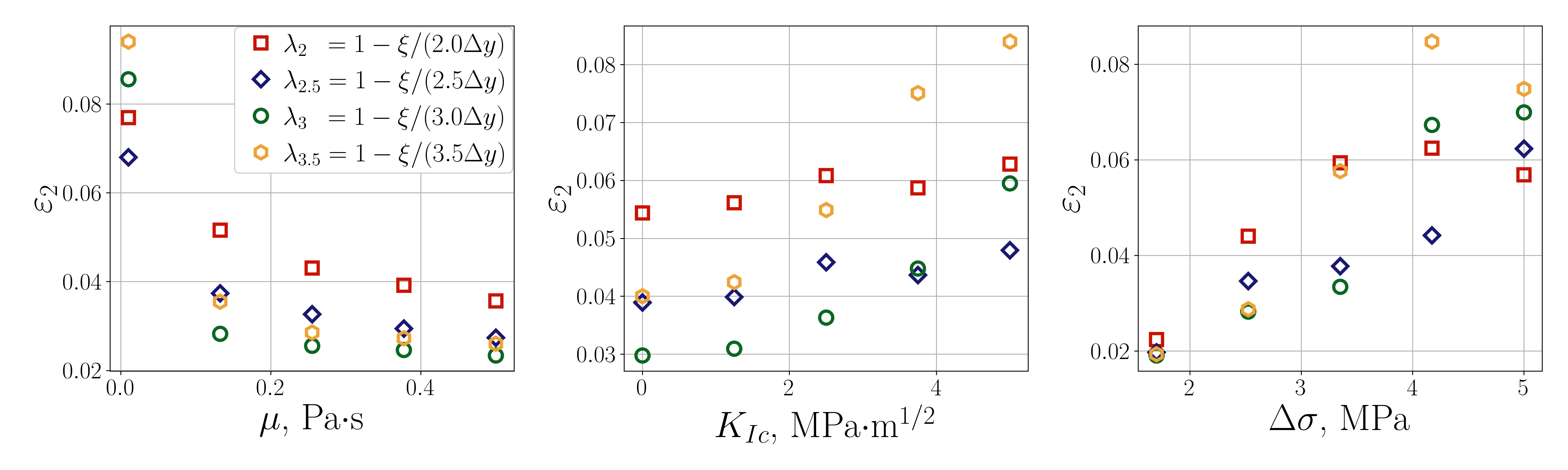}
        \caption{Comparison of the error $\varepsilon_2$ defined in~(\ref{eps2}) for different values of the stress relaxation distance $k \in \{2, 2.5, 3, 3.5\}$.}
        \label{img:stress_calibration_error_rng}
    \end{figure}

    Figure~\ref{img:stress_calibration_error_rng} shows the comparison of the error $\varepsilon_2$ defined in~(\ref{eps2}) for different values of the stress relaxation distance $k$. Different values of fluid viscosity $\mu$, fracture toughness $K_\mathrm{Ic}$, and magnitude of the stress barrier $\Delta\sigma$ are considered. For most of the parameters, the minimum of the error $\varepsilon_2$ is reached for the case $k = 3$. However, it is worth noting that the difference between the results for $k = 2$, $k = 2.5$, and $k = 3$ is not very significant. Thus, for simplicity, we choose $\lambda_3(\xi)$ as the optimal stress relaxation factor.

    \begin{figure}
        \centering
        \includegraphics[width=1.0\linewidth]{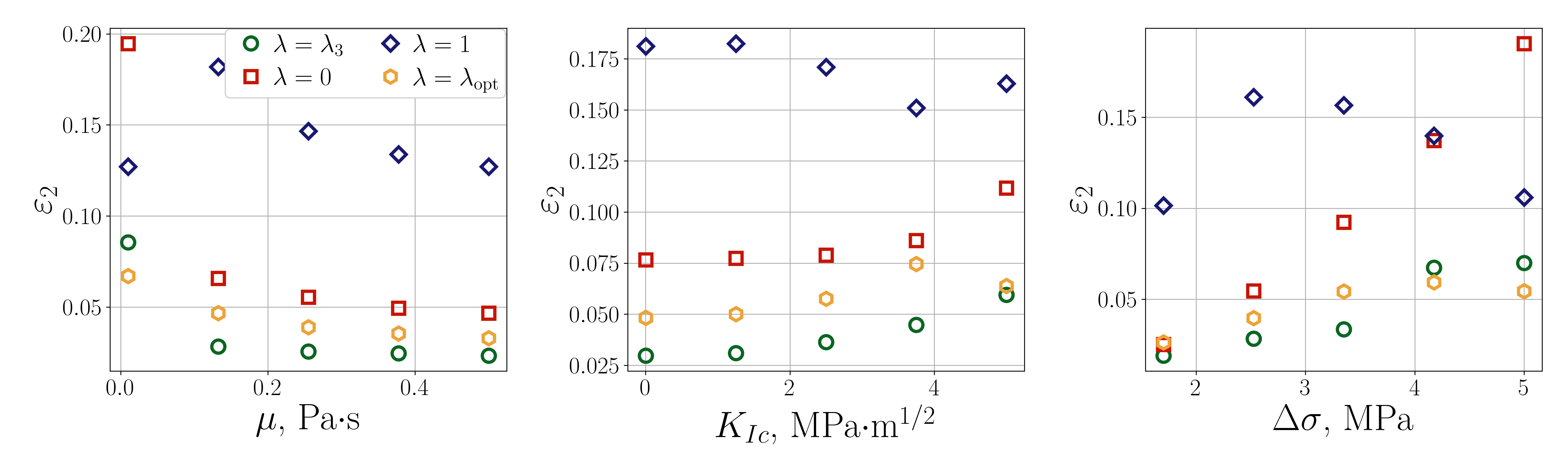}
        \caption{Comparison of the error $\varepsilon_2$ for the universal asymptote ($\lambda = 0$), the stress-corrected asymptote without relaxation ($\lambda = 1$), the stress-corrected asymptote with optimal relaxation $\lambda_3(\xi)$, and $\lambda_\mathrm{opt}$ defined in~\eqref{lambda_opt_plane_strain}.}
        \label{img:stress_calibration_error_opt}
    \end{figure}

    In addition, Figure~\ref{img:stress_calibration_error_opt} shows the comparison of the corresponding errors for $\lambda_3(\xi)$ and $\lambda_\mathrm{opt}(\xi)$ defined in~\eqref{lambda_opt_plane_strain} for the case of symmetric stress barriers. We also consider the universal asymptotic solution ($\lambda = 0$) and the stress-corrected asymptote without relaxation ($\lambda = 1$). The case with $\lambda_3(\xi)$ shows the best performance for almost the entire range of parameters, and it is close to the result calculated with $\lambda_\mathrm{opt}(\xi)$. Thus, for further calculations, we use the stress-corrected asymptote with optimal relaxation, which includes solving the initial value problem~\eqref{ode_approx_regular} with the modified stress term~\eqref{ode_stress_term_corrected} with $\lambda = \lambda_3(\xi)$.

\section{Numerical examples}\label{sec:numerical_examples}
    
    \subsection{Penny-shaped fracture}\label{sec:penny_shaped}
        To verify the numerical algorithm, we first consider the case of a penny-shaped hydraulic fracture with no fluid lag~\cite{Madyarova_Penny-shaped_2003, Dontsov_penny_shaped_2016}. The problem has four vertex or limiting solutions~\cite{Savitski_Detournay_Penny-shaped_asymptotic_2002,Detournay_Propagation_regimes_2004}: storage-viscosity ($M$), storage-toughness ($K$), leak-off-viscosity ($\tilde{M}$), and leak-off-toughness ($\tilde{K}$). This section compares the developed numerical method to the semi-analytical solution for a penny-shaped hydraulic fracture~\cite{Dontsov_penny_shaped_2016}. In order to probe the limiting regimes, we consider parameters corresponding to the validity zone of the corresponding vertex solution. The common problem parameters for all propagation regimes are the following: Young's modulus $E = 32$ GPa, and Poisson's ratio $\nu = 0.3$. The remaining parameters for the corresponding regimes are summarized in Table~\ref{tab:vertex_regimes_parameters}. The fluid is injected at a constant flow rate $Q_0 = 0.01$ m$^{3}$/s for 1000 s.
        
        \begin{table}
            \begin{center}
                \begin{tabular}{cccc}
                    \toprule
                    Regime      & $\mu$, Pa $\cdot$ s & $K_\mathrm{Ic}$, MPa $\cdot$ m$^{1/2}$ & $C_L$, m / s$^{1/2}$ \\
                    \midrule
                    $K$         & $10^{-4}$           & $3$                             & 0 \\
                    $M$         & $0.1$               & $0$                             & 0 \\
                    $\tilde{M}$ & $0.1$               & $0$                             & $5 \cdot 10^{-4}$ \\
                    $\tilde{K}$ & $10^{-4}$           & $5$                             & $1 \cdot 10^{-4}$ \\
                    \bottomrule
                \end{tabular}
                \caption{Input parameters for the case of a penny-shaped hydraulic fracture.}
                \label{tab:vertex_regimes_parameters}
            \end{center}
        \end{table}

\begin{figure}
            \centering
            \includegraphics[width=1.0\linewidth]{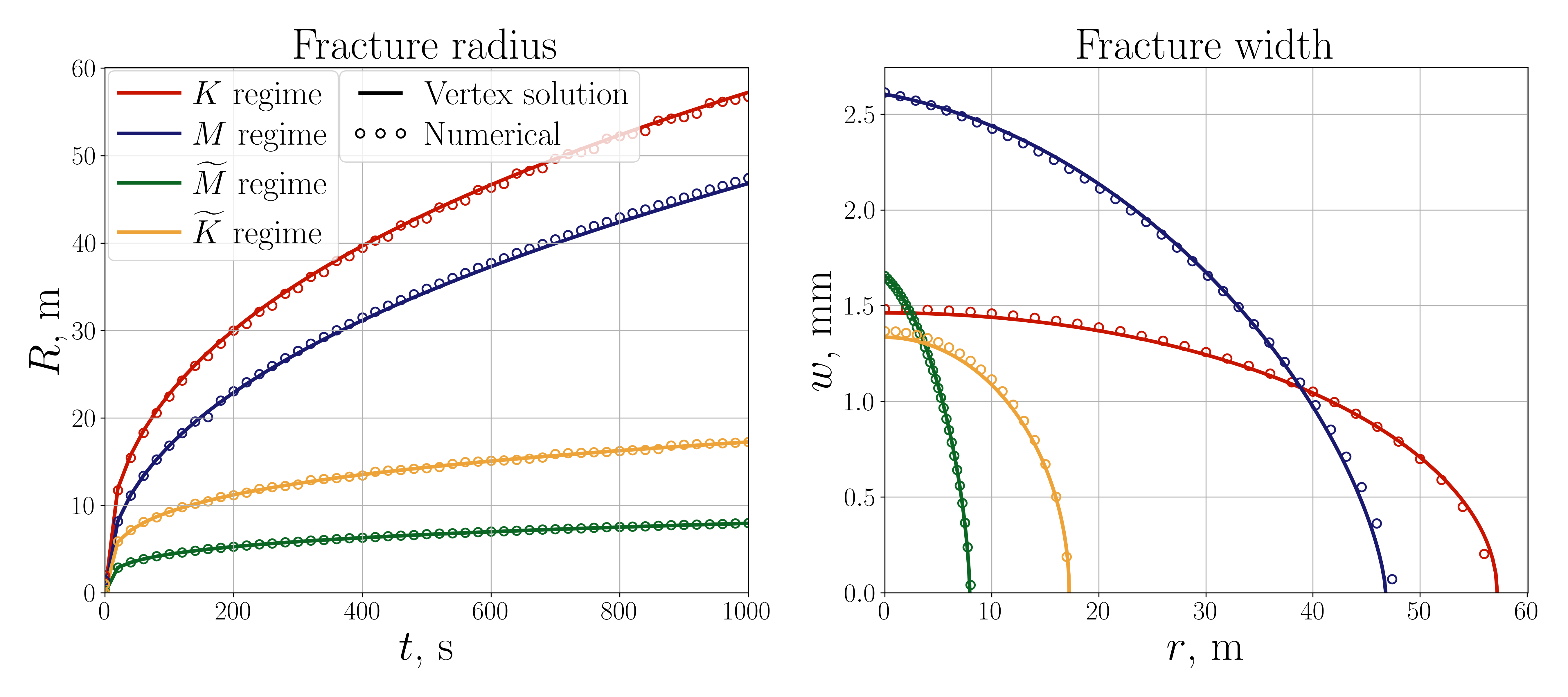}
            \caption{Results of simulations for a penny-shaped hydraulic fracture. Numerical results are shown by the circular markers, while the vertex solutions are shown by the solid lines. The left panel shows the evolution of the fracture radius, while the right panel depicts the fracture width at $t = 1000$ s.}
            \label{img:radial_comparison}
        \end{figure}

        The left panel in Figure~\ref{img:radial_comparison} shows the comparison of the fracture radius evolution for the limiting propagation regimes of the penny-shaped hydraulic fracture. Here the radius of fracture is calculated as the fracture extension along the $x$-axis at $y = 0$. For the $K$ regime, the numerical solution features step-wise behavior. This propagation pattern is associated with discretization by rectangular elements and nearly constant fluid pressure in the fracture. It is worth noting that a similar behavior of the fracture radius has been observed in~\cite{Chen_Simultaneous_growth_2019, Shovkun_Espinoza_Toughness-dominated_fracture_2019} for the toughness-dominated regime. As can be seen from the figure, the fracture radius overall closely matches the vertex solutions for all limiting regimes.
        
        The right panel in Figure~\ref{img:radial_comparison} shows the comparison of the fracture width for the different regimes at the final time $t = 1000$ s. The agreement between the numerical results and the semi-analytical solutions is excellent.
    
    \subsection{Comparison with experimental observations}
        In order to validate the numerical algorithm, we compare computations using the developed numerical algorithm with the experimental data from \cite{Bunger_Wu_PMMA_Experiment_2008}. In the experiment, the fracture propagates between two separate blocks of polymethylmethacrylate (PMMA). Hence, the fracture toughness in the contact surface is zero, and also the blocks are impermeable. For comparison with the experiment, we use the following parameters: Young's modulus $E = 3.3$ GPa, Poisson's ratio $\nu = 0.4$, fracture toughness $K_\mathrm{Ic} = 0$, and leak-off coefficient $C_L = 0$. The surface of one of the blocks is profiled so that after compression the piece-wise constant stress distribution over the contact plane is effectively imposed. The layer boundaries and the corresponding compressive stresses are:
        \begin{gather*}
            y_1 = -25 \text{ mm}, \,\, y_2 = 25 \text{ mm}, \quad \sigma_h^1 = 11.2 \text{ MPa}, \,\, \sigma_h^2 = 7 \text{ MPa}, \,\, \sigma_h^3 = 5 \text{ MPa}.
        \end{gather*}
        A fluid with viscosity $\mu = 30$ Pa $\cdot$ s is pumped with a piece-wise constant flow rate
        \begin{equation*}
            Q(t) = 
            \begin{cases}
                0.9 \cdot 10^{-9} \text{ m}^3/\text{s}, \quad &0 \text{ s} < t \leqslant 31 \text{ s}, \\
                6.5 \cdot 10^{-9} \text{ m}^3/\text{s}, \quad &31 \text{ s} < t \leqslant 151 \text{ s}, \\
                2.3 \cdot 10^{-9} \text{ m}^3/\text{s}, \quad &151 \text{ s} < t.
            \end{cases}
        \end{equation*}
        We refer to~\cite{Bunger_Wu_PMMA_Experiment_2008} for a more detailed description of the experimental procedure.

\begin{figure}
            \centering
            \includegraphics[width=1.0\linewidth]{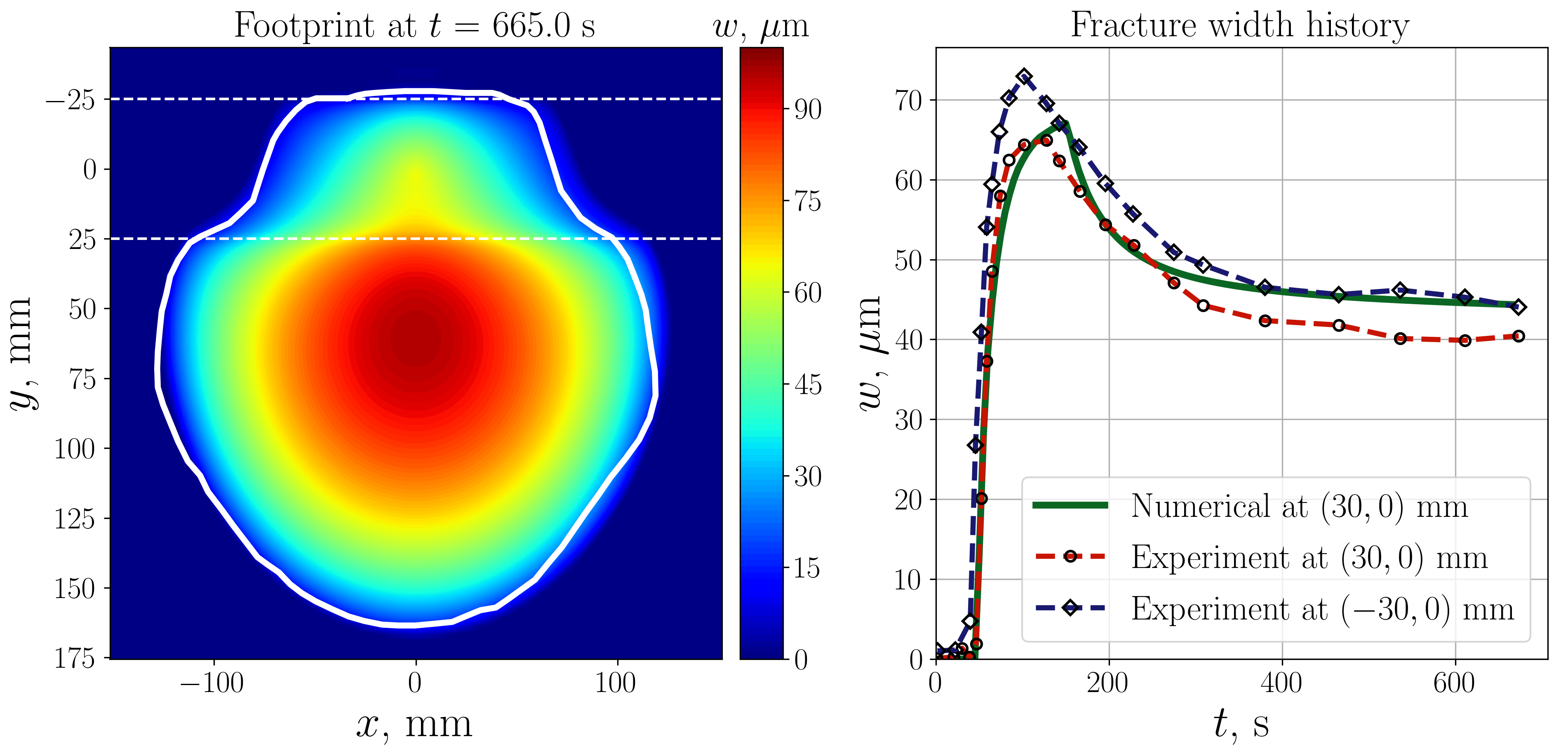}
            \caption{Left panel: numerically computed fracture width (color filling) and the fracture footprint obtained in the experiment (solid white line). Right panel: temporal evolution of the fracture width. The dashed red and blue lines depict the experimental measurements at $(30, 0)$ mm and $(-30, 0)$ mm, respectively. The solid green line shows the numerical result.}
            \label{img:bunger_comparison}
        \end{figure}

        The left panel in Figure~\ref{img:bunger_comparison} shows the numerical results for the fracture width $w$ at $t = 665$ s. The solid white line shows the fracture footprint obtained from the experimental data. The horizontal dashed white lines depict the boundaries of the stress layers. The comparison shows a good agreement between the numerical results and experimental data for the fracture footprint. The right panel in Figure~\ref{img:bunger_comparison} displays the comparison of the fracture width evolution near the injection point. In the experiment, the fracture width is measured at $(30, 0)$ mm and $(-30, 0)$ mm. The fracture width computed numerically is evaluated only at $(30, 0)$ mm due to symmetry. As can be seen from Figure~\ref{img:bunger_comparison}, the numerical calculations of the fracture width closely match the data measured in the experiment within the measurement accuracy.
        
    \subsection{Fracture closure}
        The next example illustrates the fracture closure algorithm. In particular, the three-layered geometry of compressive stress $\sigma_h$ and the leak-off coefficient $C_L$ is considered, such that there are two disconnected fracture segments after closure. The problem parameters are the following: Young's modulus $E = 9.5$ GPa, Poisson's ratio $\nu = 0.2$, fracture toughness $K_\mathrm{Ic} = 1$ MPa$\cdot$m$^{1/2}$, and fluid viscosity $\mu = 0.1$ Pa$\cdot$s. The central layer is surrounded by two layers with lower compressive stress and zero leak-off coefficient:
        \begin{equation*}
            \sigma_h^1 = \sigma_h^3 = 6.5 \text{ MPa}, \,\, \sigma_h^2 = 7.5 \text{ MPa}, \quad C_L^1 = C_L^3 = 0, \,\, C_L^2 = 5 \cdot 10^{-5} \text{ m/s}^{1/2}.
        \end{equation*}
        The lower value of stress in the surrounding layers allows the fracture front to grow quicker once the boundary of the layers is reached. Fluid is injected in the middle of the central layer with the piece-wise constant flow rate
        \begin{equation*}
            Q(t) = 
            \begin{cases}
                0.02 \text{ m}^3/\text{s}, \quad &0   \text{ s} < t \leqslant 550 \text{ s}, \\
                0    \text{ m}^3/\text{s}, \quad &550 \text{ s} < t.
            \end{cases}
        \end{equation*}

        \begin{figure}
            \centering
            \includegraphics[width=1.0\linewidth]{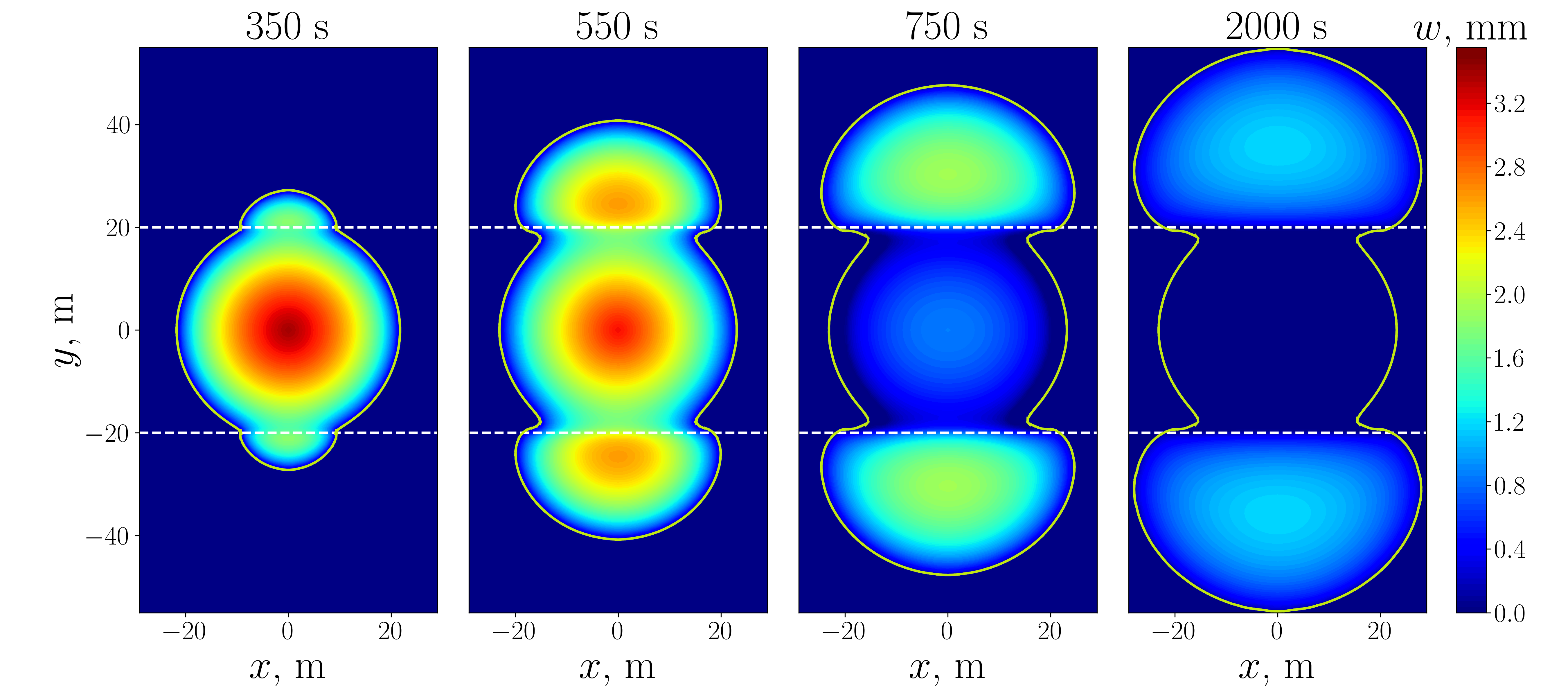}
            \caption{The fracture width distribution for the example of fracture closure evaluated at $t \in \{350, 550, 750, 2000\}$ s. The horizontal dashed lines show layer boundaries.}
            \label{img:unconnected_width}
        \end{figure}

        Figure~\ref{img:unconnected_width} illustrates the calculated fracture width for different time instants. The fracture front reaches the boundaries of the central layer and starts to grow quicker due to the lower value of stress at $t = 350$ s. Fluid injection stops at $t = 550$ s. Then, the fracture in the central layer starts to close due to fluid leak-off into the reservoir. At the same time, the fracture portions in the top and bottom layers continue to propagate because the leak-off coefficient in these layers is equal to zero. At $t = 2000$ s, the fracture in the central layer is completely closed, and the contact condition $w(x, y) = \wmin(x, y)$ is satisfied. The two fracture regions become separated by the zone where fracture walls are in contact.

\section{Mesh sensitivity}\label{sec:mesh_sensitivity}
    In this section, we investigate the sensitivity of the hydraulic fracture geometry to mesh size. The mesh that is used for all cases is square, i.e. $\Delta x = \Delta y$. An adaptive time stepping is applied to ensure convergence of the numerical algorithm. As was stated at the beginning of this paper, the primary purpose of using the stress-corrected asymptote is to improve accuracy for coarse meshes, which in turn allows using such coarse meshes for practical problems to obtain the numerical result quicker.

    \subsection{Constant height hydraulic fracture}\label{sec:pkn_fracture}
        We first consider the propagation of a height-contained fracture or Perkins\--Kern\--Nordgren (PKN) fracture~\cite{Perkins_PKN_Fracture_1961, Nordgren_PKN_1972} in which two layers with high stress surround the reservoir and restrict vertical height growth. The problem of PKN fracture has four limiting regimes of propagation~\cite{Dontsov_Analysis_PKN_2022, Economides_Reservoir_Stimulation_1989} that are determined by the dominance of fluid viscosity or rock toughness as well as a feature either small or large fluid leak-off in the surrounding rock. In this section, we consider only the storage toughness ($K$) and storage viscosity ($M$) regimes corresponding to no leak-off.
        
        To check the sensitivity of the fracture geometry to mesh size, we perform simulations for a different number of elements $n_c$ in the central layer. The height of the reservoir layer is $H = 20$ m, and the surrounding layers have compressive stress $\sigma_h^1 = \sigma_h^3 = 30$ MPa, while the compressive stress in the middle layer is $\sigma_h^2 = 20$ MPa. The parameters corresponding to the $K$ and $M$ regimes are outlined in Table~\ref{tab:pkn_regimes_params}. The simulations are performed for 700 seconds of fluid injection with a constant flow rate.

        \begin{table}
            \begin{center}
                \begin{tabular}{ccccccc}
                    \toprule
                    Regime      & $E$, GPa & $\nu$ & $K_\mathrm{Ic}$, MPa$\cdot$m$^{1/2}$ & $C_L$, m/s$^{1/2}$ & $\mu$, Pa$\cdot$s & $Q$, m$^3$/s \\
                    \midrule
                    $K$         & 1        & 0.3   & 1.5                                  & 0                  & 0.01              & 0.05      \\
                    $M$         & 3        & 0.3   & 0.5                                  & 0                  & 0.1               & 0.02      \\
                    \bottomrule
                \end{tabular}
                \caption{Parameters for the storage toughness ($K$) and storage viscosity ($M$) regimes of a constant height hydraulic fracture.}
                \label{tab:pkn_regimes_params}
            \end{center}
        \end{table}

        \begin{figure}
            \centering
            \includegraphics[width=1.0\linewidth]{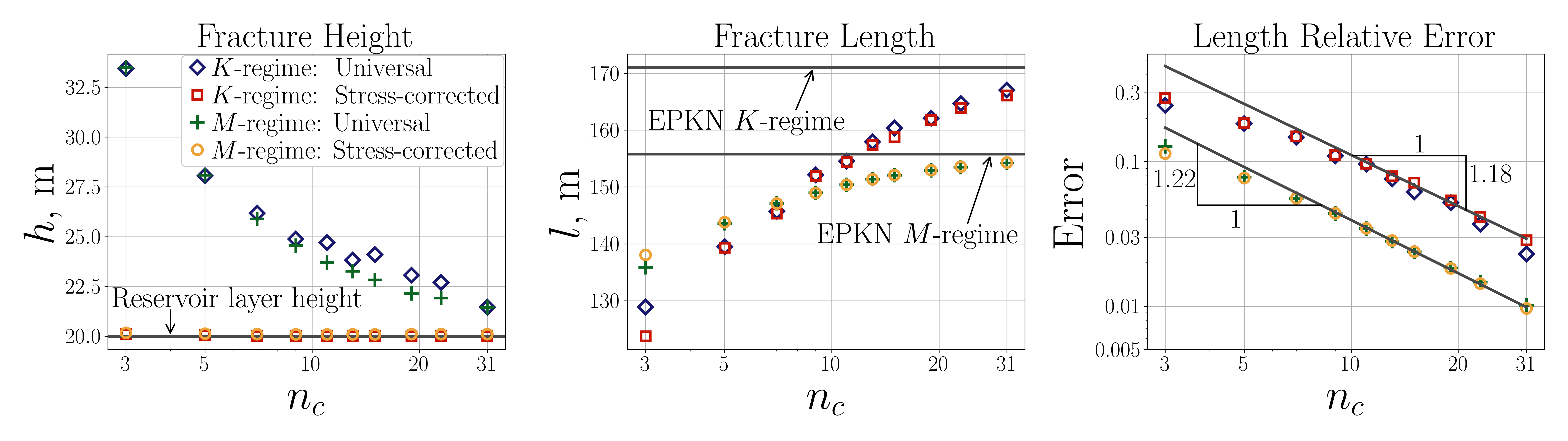}
            \caption{Fracture height (left), fracture length (center), and the error of the fracture length compared to the reference solution (right) for a constant height fracture versus the number of elements $n_c$ in the central layer. Blue diamonds correspond to $K$-regime and the universal asymptote, red circles to $K$-regime and the stress-corrected asymptote, green squares to the $M$-regime and the universal asymptote, and yellow hexagons to the $M$-regime and the stress-corrected asymptote. Gray horizontal lines denote the reference length computed by EPKN model~\cite{Dontsov_Peirce_EPKN_2016}.}
            \label{img:pkn_convergence}
        \end{figure}

        The left panel in Figure~\ref{img:pkn_convergence} plots the fracture height versus the number of elements $n_c$ in the reservoir layer. Blue diamonds correspond to the $K$-regime and the universal asymptote, red circles to $K$-regime and the stress-corrected asymptote, green squares to the $M$-regime and the universal asymptote, and yellow hexagons to the $M$-regime and the stress-corrected asymptote. The gray horizontal line shows the height of the reservoir layer. In the case of the universal asymptotic solution, the discrepancy between the fracture height and the reservoir layer height is proportional to double the element height, i.e. $2\Delta y$, for both $K$ and $M$ regimes. This error corresponds to the growth of approximately one element beyond the layer boundary and is an unwanted feature of the original ILSA implementation. In the case of stress-corrected asymptote, the fracture height almost precisely coincides with the reservoir layer height for both of the regimes and all meshes.

        The central panel in Figure~\ref{img:pkn_convergence} shows the final fracture length versus the number of elements $n_c$ in the reservoir layer. The gray horizontal lines signify the reference fracture length computed by EPKN model with non-local elasticity~\cite{Dontsov_Peirce_EPKN_2016}. For the $K$-regime, both the universal asymptotic solution and the stress-corrected asymptote predict a considerably smaller length than the reference solution. The discrepancy in length between the coarsest mesh ($n_c = 3$) and the finest mesh ($n_c = 31$) is approximately 25 percent. For the $M$-regime, this difference reduces to approximately 10 percent. It is interesting to observe that the improvement in the height calculation does not noticeably impact the fracture length. The reason for this lies in the fact that the elements that are located outside the central layer are practically closed and do not have any fluid volume. Therefore, the fracture volume in each vertical cross-section is approximately the same no matter what asymptote is used, which ultimately results in similar fracture length predictions.
        
        The right panel in Figure~\ref{img:pkn_convergence} depicts the error of the fracture length relative to the reference solution. One can observe a convergence rate of approximately 1.2 for both $K$ and $M$ regimes. It is worth noting that the fracture length converges at the same rate for both the universal asymptote and the stress-corrected asymptote.

        It is important to note that a similar issue of length convergence for a constant height fracture was mentioned in Dontsov et al.~\cite{Dontsov_Front_Tracking_Comparison_2022}. In the latter work, inaccurate computation of the fracture height was cited as the primary reason for length error. However, in the case of the stress-corrected asymptote, the fracture growth is arrested exactly at the location of the stress barrier, but the length accuracy remains practically unchanged. One possible explanation is that the asymptotic solution becomes less accurate for high fracture front curvatures, where the validity zone of the plane strain assumption for the tip asymptote is reduced. This problem requires further investigation and is beyond the scope of this study.

    \subsection{Symmetric stress barriers}
        This section examines the sensitivity of the fracture geometry to mesh size for the case of moderate symmetric stress barriers, i.e. when some fracture height growth is permitted. Similarly to the previous section, we consider the two sets of parameters presented in Table~\ref{tab:pkn_regimes_params}. To promote the fracture height growth, we reduce the compressive stress in the surrounding layers to $\sigma_h^1 = \sigma_h^3 = 20.15$ MPa for the $K$ regime and to $\sigma_h^1 = \sigma_h^3 = 20.35$ MPa for the $M$ regime. The confining stress in the central layer remains the same, i.e. $\sigma_h^2 = 20$ MPa.

        \begin{figure}
            \centering
            \includegraphics[width=1.0\linewidth]{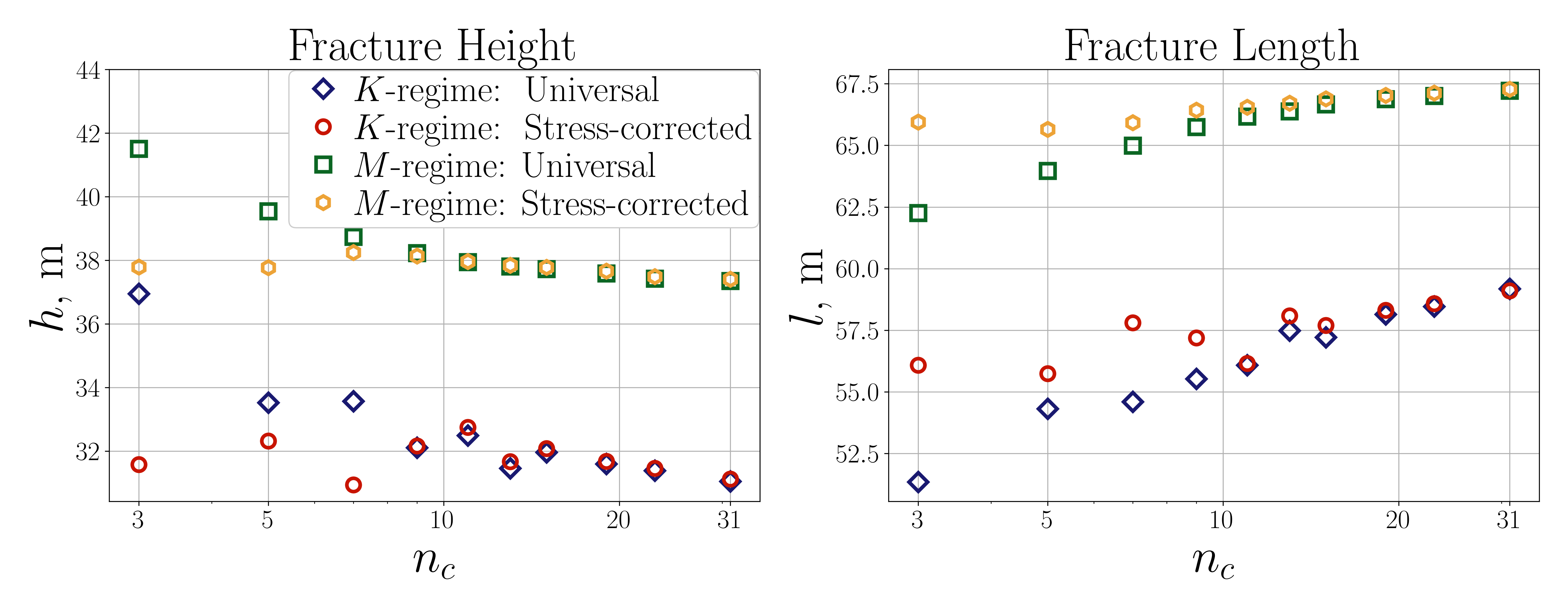}
            \caption{Fracture height (left) and fracture length (right) at $t = 300$ seconds for a symmetric stress barrier versus the number of elements $n_c$ in the central layer. Blue diamonds correspond to the $K$ regime and the universal asymptote, red circles to the $K$ regime and the stress-corrected asymptote, green squares to the $M$ regime and the universal asymptote, and yellow hexagons to the $M$ regime and the stress-corrected asymptote. }
            \label{img:p3d_mesh_convergence}
        \end{figure}

        Figure~\ref{img:p3d_mesh_convergence} shows the fracture height (left) and fracture length (right) versus the number of elements $n_c$ in the reservoir layer. As for the constant height case, the blue diamonds correspond to the $K$ regime and the universal asymptote, red circles to the $K$ regime and the stress-corrected asymptote, green squares to the $M$ regime and the universal asymptote, and yellow hexagons to the $M$ regime and the stress-corrected asymptote. The non-monotonic behavior of the fracture height and length for the $K$ regime is related to the step-wise fracture propagation, as was discussed in Section~\ref{sec:penny_shaped}. The universal asymptote considerably overestimates the fracture height on a coarse grid, which also leads to significant errors in length since the fracture volume remains the same. The stress-corrected asymptote, on the other hand, estimates the fracture height more accurately, resulting in accurate computation of the fracture length as well.

        \begin{figure}
            \begin{subfigure}[b]{0.99\textwidth}
                \centering
                \includegraphics[width=1.0\linewidth]{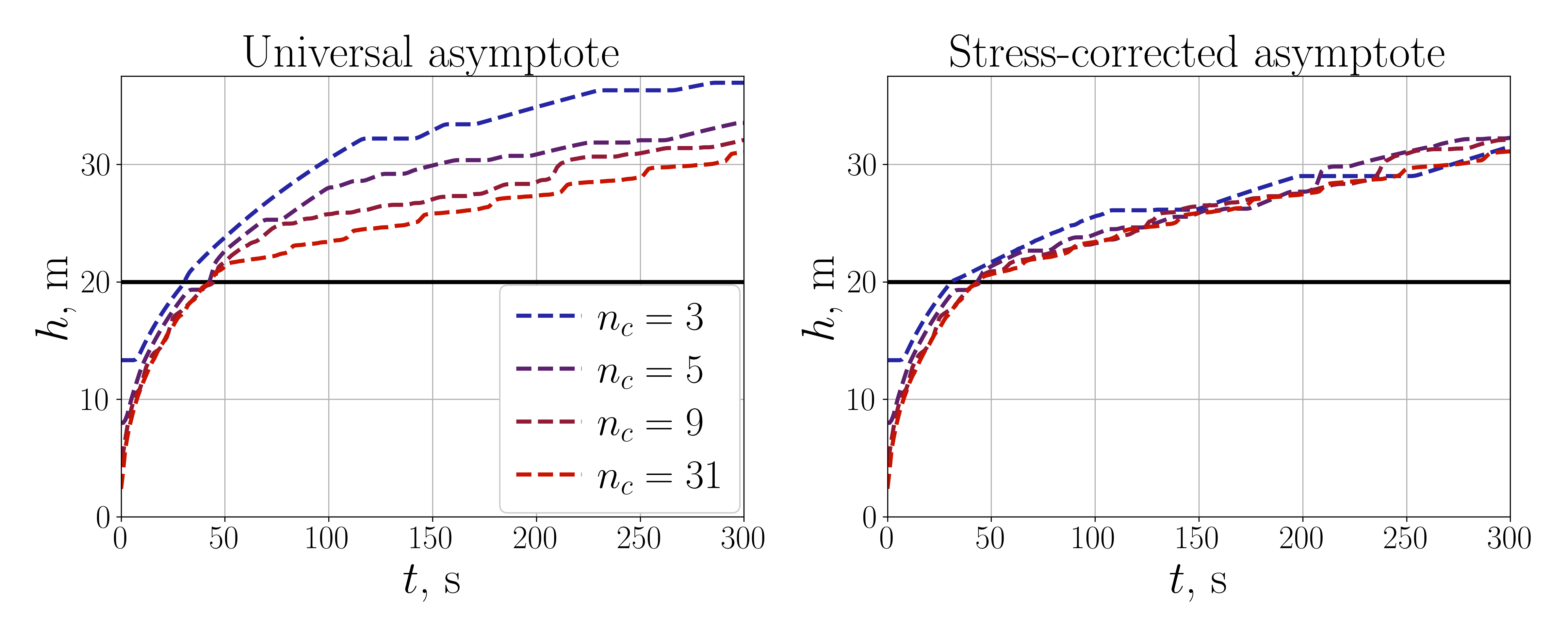}
                \caption{$K$ regime.}
            \end{subfigure}
            \\
            \begin{subfigure}[b]{0.99\textwidth}
                \centering
                \includegraphics[width=1.0\linewidth]{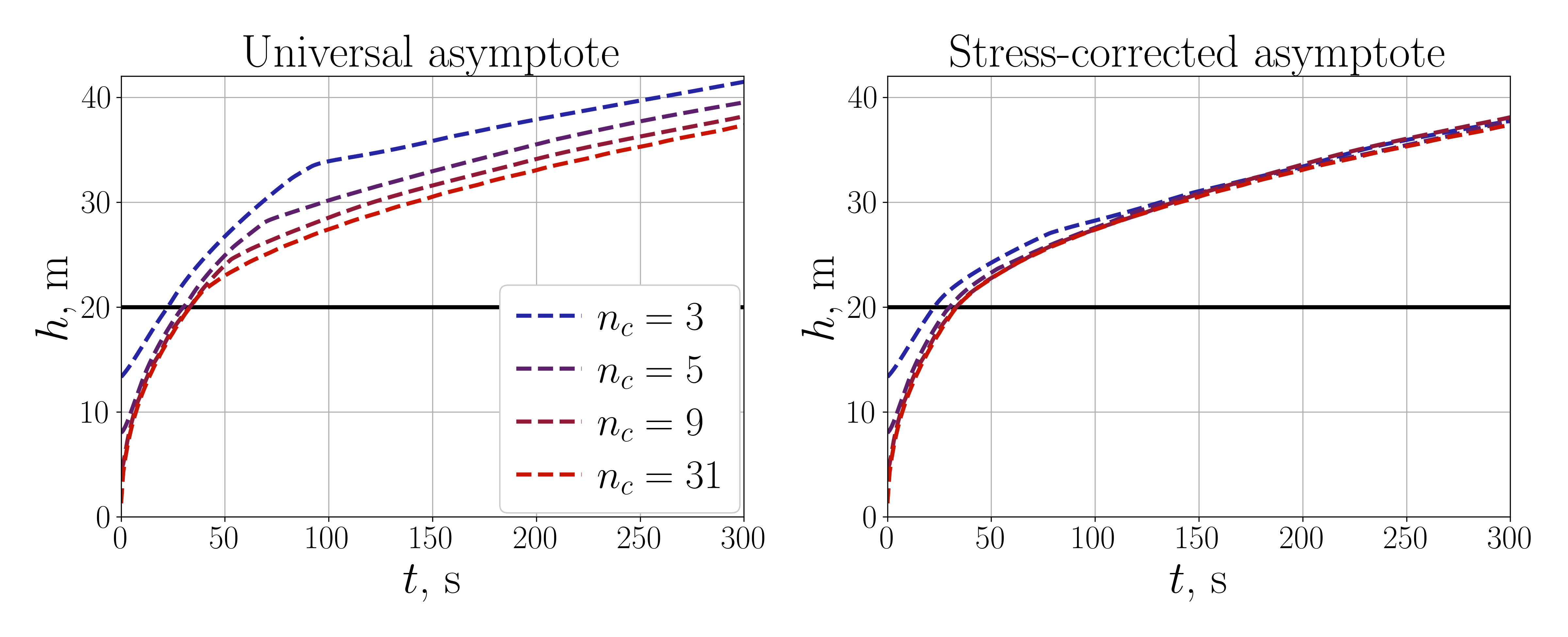}
                \caption{$M$ regime.}
            \end{subfigure}
            \caption{Fracture height versus time for $K$ regime $(a)$ and $M$ regime $(b)$. The left panel for each row shows the fracture height for the universal asymptote, while the right panel depicts the result for the stress-corrected asymptote. The black horizontal line shows the height of the reservoir layer, while $n_c$ denotes the number of elements in the central layer.}
            \label{img:p3d_height_zoom}
        \end{figure}

        Figure~\ref{img:p3d_height_zoom} shows the fracture height versus time for the $K$ regime $(a)$ and the $M$ regime $(b)$. The left panel for each row shows the fracture height for the universal asymptote, while the right panel shows the results for the stress-corrected asymptote. The black horizontal lines signify the height of the reservoir layer. For the case of the universal asymptote, the fracture height prediction on a coarse mesh has a significant error relative to the fine mesh result. In contrast with the universal asymptote, the stress-corrected asymptote provides a more accurate prediction of layer crossing. The fracture height practically does not depend on the mesh size when the fracture crosses the layer boundary. Note that the $K$ regime case features oscillatory behavior, as was also previously observed for the radial fracture case.

    \subsection{Thin high stress layer}
        This section aims to examine the fracture propagation for the case of a configuration with a thin layer that has high compressive stress. The parameters used in the simulation are the following: Young's modulus $E = 20$ GPa, Poisson's ratio $\nu = 0.3$, fracture toughness $K_\mathrm{Ic} = 0.5$ MPa$\cdot$m$^{1/2}$, leak-off coefficient $C_L = 0$, fluid viscosity $\mu = 0.1$ Pa$\cdot$s, and injection flowrate $Q = 0.1$ m$^3$/s. The layer boundaries and compressive stresses are defined as follows:
        \begin{gather*}
            y_1 = -7.5 \text{ m}, \,\, y_2 = 7.5 \text{ m}, \,\, y_3 = 12.5 \text{ m}, \,\, y_4 = 37.5 \text{ m}, \\
            \sigma_h^1 = 35 \text{ MPa}, \,\, \sigma_h^2 = 26 \text{ MPa}, \,\, \sigma_h^3 = 30 \text{ MPa}, \,\, \sigma_h^4 = 26 \text{ MPa}, \,\, \sigma_h^5 = 35 \text{ MPa}.
        \end{gather*}

        The thin layer is located between $y_2 = 7.5$ m and $y_3 = 12.5$ m and is subject to the confining stress of $\sigma_h^3 = 30$ MPa. The computations are performed for different numbers of elements in the thin layer $n \in \{1, 2, 3, 5, 10\}$, and the mesh is such that $\Delta x = \Delta y = (y_3 - y_2) / n$. 
        
\begin{figure}
            \centering
            \includegraphics[width=1.0\linewidth]{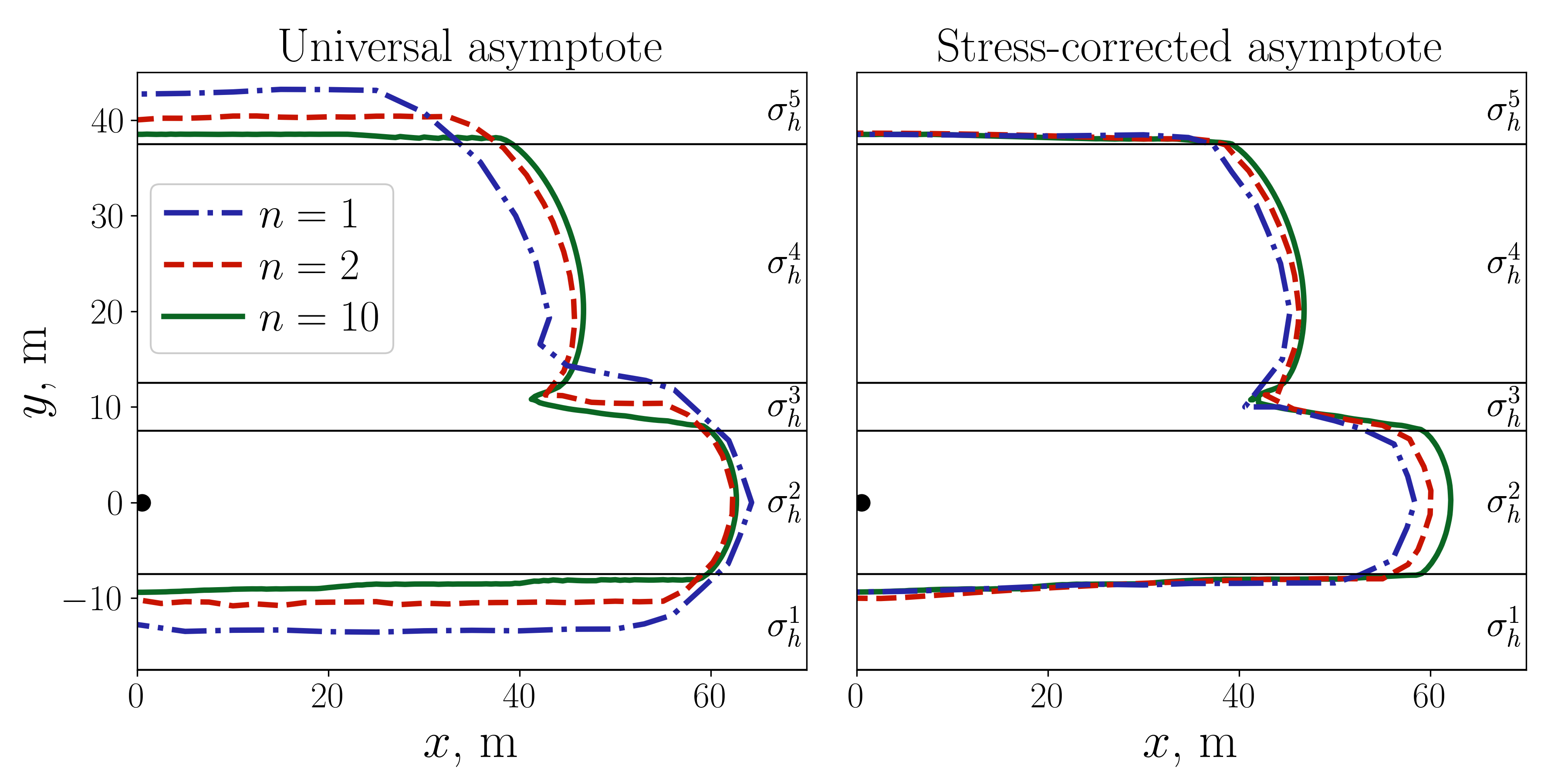}
            \caption{Fracture footprint for the case of thin high stress layer at $t = 150$ seconds for the universal asymptote (left) and the stress-corrected asymptote (right). The number of elements in the thin layer is denoted by $n$. The horizontal black lines show the layer boundaries, while the black circular marker highlights the injection point.}
            \label{img:thin_layer_width}
        \end{figure}

        Figure~\ref{img:thin_layer_width} shows the fracture footprint at $t = 150$ seconds for the universal asymptote (left) and the stress-corrected asymptote (right). The blue dash-dotted lines depict the footprint for coarse mesh with $n = 1$, the red dashed lines show the result for mesh with $n = 2$, and the solid green lines correspond to the footprint for fine mesh with $n = 10$. As can be seen from the figure, the stress-corrected  asymptote gives a very accurate height approximation, which is in contrast to the universal asymptote. The reduced accuracy of the fracture height prediction for the universal asymptote is related to the issue illustrated on the left panel in Figure~\ref{img:pkn_convergence}. Since the tip elements don't account for stress layers, the fracture always grows at least one element beyond the layer boundary. The fracture length depends on the mesh size for both the universal asymptote and the stress-corrected asymptote. However, the length discrepancy between the coarse and fine meshes does not exceed 10 percent for the universal asymptote and 6 percent for the stress-corrected asymptote, which is acceptable for most practical cases. Note that such a dependence of length on the mesh size can be related to the fact that the fracture in the layers above and below the thin layer resembles PKN fracture, for which a relatively strong length discrepancy is observed, see discussion in section~\ref{sec:pkn_fracture}.

\begin{figure}
            \centering
            \includegraphics[width=1.0\linewidth]{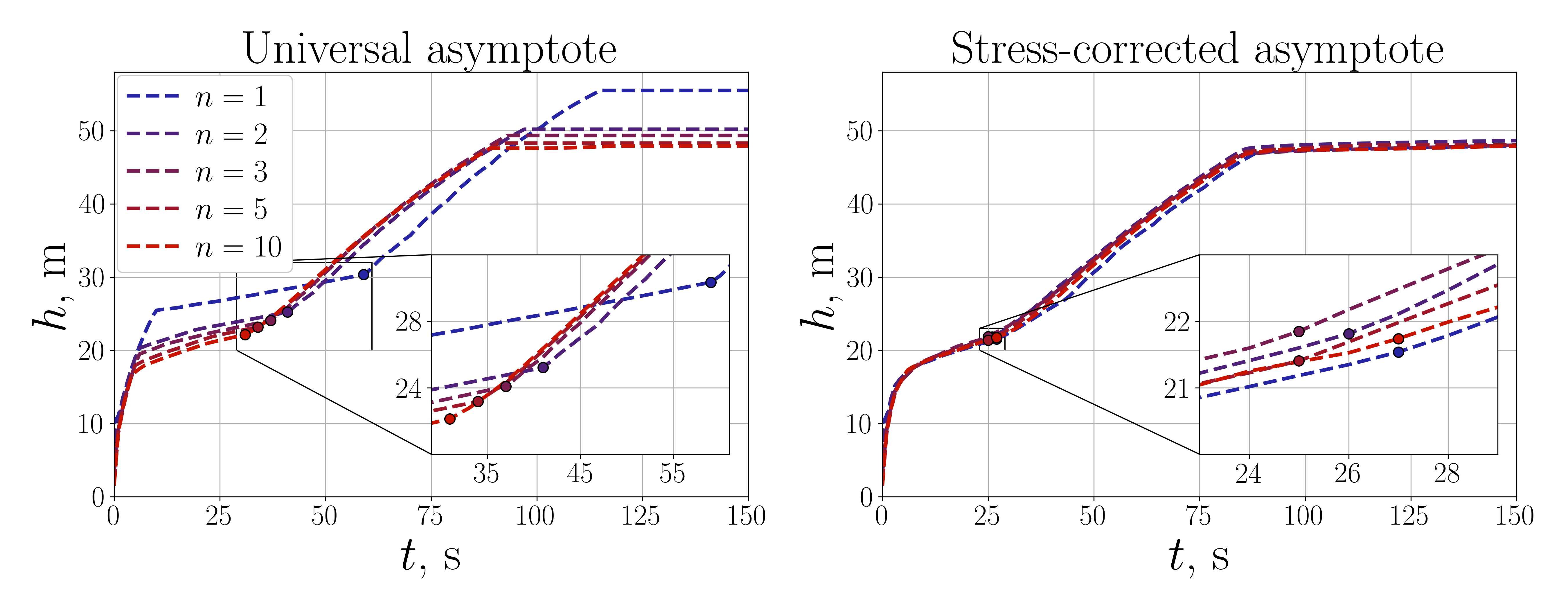}
            \caption{Fracture height versus time for the case with high stress thin layer. Different lines correspond to different numbers of elements $n \in \{1, 2, 3, 5, 10\}$ in the thin layer. The left panel shows the results of the universal asymptote, while the results of the stress-corrected asymptote are shown in the right picture. Circular markers indicate the crossing time of the thin layer. }
            \label{img:thin_layer_height}
        \end{figure}

        In order to quantify accuracy associated with fracture crossing the thin layer, Figure~\ref{img:thin_layer_height} plots the fracture height versus time. The left panel corresponds to the universal asymptote, while the results for the stress-corrected asymptote are shown on the right. Different numbers of elements in the thin layer $n \in \{1, 2, 3, 5, 10\}$ are considered. Circular markers indicate the time instances corresponding to the crossing of the thin layer. In the case of the stress-corrected asymptote, the crossing occurs when a tip element is created above the thin layer. For the universal asymptote, the thin layer is considered crossed when the survey element appears above the thin layer since the universal asymptote does not ``feel'' the stress layer. As can be seen from the figure, for the case of universal asymptote, the time of crossing varies significantly versus mesh. On the other hand, in the case of the stress-corrected asymptote, the evolution of fracture height and the thin layer crossing time are practically independent of the mesh size. Thus, the stress-corrected asymptote can significantly increase the accuracy of simulations when thin layers are present in the model.

\section{Conclusions}
    This paper presents a fracture front tracking algorithm for an implicit level set algorithm (ILSA) that accounts for the effect of multiple stress layers. The front tracking procedure employs the stress-corrected asymptotic solution that accounts for the effect of stress layers in the multiscale tip asymptote~\cite{Valov_stress_corrected_asymptote_2023}. To apply the latter stress-corrected asymptote, the core components of the original ILSA approach, such as locating the fracture front and tip volume calculation, have been extended to include the effect of multiple horizontal stress layers. Additionally, we have determined the stress relaxation factor that does not depend on the geometry of the reservoir layers to increase the validity region of the stress-corrected asymptote in the case of a finite planar fracture. These modifications significantly increase the accuracy of ILSA algorithm for a layered formation, especially in the case of a coarse computational mesh. The accuracy of the presented solver is verified by comparing the results to reference numerical calculations for a radial hydraulic fracture in different propagation regimes and experimental results conducted for a transparent PMMA block. In order to illustrate the fracture closure algorithm, we consider the case of three-layered geometry, such that there are two disconnected segments after closure.

    In addition, we have investigated the sensitivity of fracture geometry to mesh size since the primary goal of using the stress-corrected asymptote is to improve accuracy for coarse meshes. For a height-contained fracture, the height error corresponds to the growth of approximately one element outside the layer boundary for the case of the universal asymptote. While in the algorithm with the stress-corrected asymptote, the fracture growth is arrested almost precisely at the location of the stress barrier for all meshes. Additionally, both algorithms with the universal asymptote and the stress-corrected asymptote predict a notably smaller length than the reference solution. This discrepancy can be related to the reduction of accuracy of the asymptotic solution for high fracture front curvatures. In the case of a moderate stress barrier, the stress-corrected asymptote provides a more accurate prediction of height and length comparing the universal asymptote. Furthermore, the accuracy of the developed ILSA algorithm with the stress-corrected asymptote has been examined for the case of a thin layer with high confining stress. It is concluded that the developed algorithm provides a very accurate height prediction for all meshes, which is in contrast to the standard ILSA implementation with the universal asymptotic solution that shows strong mesh sensitivity.
        
\appendix

\section{Discretized governing equations}\label{appendix:discretization}
    In order to obtain discretized equations, a piece-wise constant approximation of the fracture width $w(x, y, t)$ and the fluid pressure $p(x, y, t)$ is employed. It is assumed that the fracture width $w$, the fluid pressure $p$, and the fracture footprint are known at the previous time step $t - \Delta t$.

    Using the displacement discontinuity method~\cite{Crouch_Starfield_BEM_1983} and assuming a piece-wise constant fracture width approximation, the elasticity equation~\eqref{elasticity_equation} in the case of non-constrained fracture ($w(x, y, t) > \wmin(x, y, t)$ and $T_n = p$) takes the following form
    \begin{equation}\label{elasticity_equation_discrete}
        p_{i, j}(t) = {\sigma_h}_{i, j} + \sum_{k, \, l} C_{i, j; k, l} \, w_{k, l}(t),
    \end{equation}
    where $C_{i, j; k, l}$ are the coefficients of the elasticity matrix representing the influence of element $(k, l)$ on element $(i, j)$. For the case of a homogeneous linear elastic medium, as considered in this study, the elasticity matrix can be computed as~\cite{Peirce_Detournay_ILSA_2008}
    \begin{equation}\label{elasticity_matrix_homogeneous}
        C_{i, j; k, l} = -\frac{E'}{8\pi} \left[ \frac{\sqrt{(x_{i} - x)^2 + (y_{j} - y)^2}}{(x_{i} - x)(y_{j} - y)} \right]_{x = x_{k} - \Delta x/2, ~ y = y_{l} - \Delta y/2}^{x = x_{k} + \Delta x/2, ~ y = y_{l} + \Delta y/2},
    \end{equation}
    where $\left[f\right]_{x = x_1, y = y_1}^{x = x_2, y = y_2} = f(x_1, y_1) + f(x_2, y_2) - f(x_1, y_2) - f(x_2, y_1)$.

    The lubrication equation~\eqref{reynolds_equation} is discretized using the finite volume method, as in~\cite{Dontsov_Peirce_ILSA_2017}. By using backward Euler's scheme for the time integration from $t - \Delta t$ to $t$, we obtain the following system of equations
    \begin{equation}\label{reynolds_equation_discrete}
        w_{i, j}(t) - w_{i, j}(t - \Delta t) = \Delta t \left[A p\right]_{i, j} + \Delta t \, Q_{i, j} - \Delta \mathcal{L}_{i, j},
    \end{equation}
    where $Q_{i, j}$ denotes the source term, $\Delta \mathcal{L}_{i, j}$ captures the fluid leak-off for the element $(i, j)$ over time $[t - \Delta t, t]$. For the channel elements, the leak-off term can be calculated as
    \begin{equation}\label{leakoff_discrete_channel}
        \Delta \mathcal{L}_{i, j} = 2 \Cprime \left( \sqrt{t - t_{0_{i, j}}\vphantom{\Delta}} - \sqrt{t - \Delta t - t_{0_{i, j}}} \right),
    \end{equation}
    where $t_{0_{i, j}}$ is the time instance at which the fracture front was located at the center of the channel element $(x_i, y_j)$. The fluid fluxes across the element edges are calculated using the central difference
    \begin{eqnarray}\label{flux_matrix}
        \left[ Ap \right]_{i, j} &=& \frac{1}{\Delta x} \left[ M_{i + \frac{1}{2}, j} \frac{p_{i + 1, j} - p_{i, j}}{\Delta x} - M_{i - \frac{1}{2}, j} \frac{p_{i, j} - p_{i - 1, j}}{\Delta x} \right] \notag \\
        &+& \frac{1}{\Delta y} \left[ M_{i, j + \frac{1}{2}} \frac{p_{i, j + 1} - p_{i, j}}{\Delta y} - M_{i, j - \frac{1}{2}} \frac{p_{i, j} - p_{i, j - 1}}{\Delta y} \right],
    \end{eqnarray}
    where the corresponding fluid mobilities are defined as
    \begin{equation}\label{fluid_modility_flux}
        M_{i \pm \frac{1}{2}, j} = \frac{1}{2}\left(\frac{w_{i \pm 1, j}^3}{\muprime_{i \pm 1, j}} + \frac{w_{i, j}^3}{\muprime_{i, j}}\right).
    \end{equation}
    
    In order to incorporate the asymptotic solution into the front tracking algorithm it is convenient to split variables into the channel and tip values. In the following expressions, we will use the superscript ``$c$'' for the channel variables and the superscript ``$t$'' for the tip variables.
    
    The elasticity matrix $\bbC$ is calculated using equation~\eqref{elasticity_matrix_homogeneous}, the flux matrix $A$ in the equation~\eqref{flux_matrix} is computed as $\bbA = A(\bw)$. The tip widths $\bw^t$ are determined by integrating the asymptotic solution. Thus, in the tip elements, the fluid pressure vector $\bp^t$ is the primary unknown. In the channel elements, the fluid pressure vector $\bp^c$ is determined using  equation~\eqref{elasticity_equation_discrete} and the fracture width vector $\bw^c$ is unknown. Substituting~\eqref{elasticity_equation_discrete} for the channel elements in the equation~\eqref{reynolds_equation_discrete} we obtain the following system of equations
    \begin{equation}\label{slae_without_contact}
        \begin{bmatrix*}[r]
            \bbI - \Delta t \bbA^{cc} \bbC^{cc} & -\Delta t \bbA^{ct} \\
            -\Delta t \bbA^{tc} \bbC^{cc}       & -\Delta t \bbA^{tt}
        \end{bmatrix*}
        \begin{bmatrix}
            \bw^c \\
            \bp^t
        \end{bmatrix}
        =
        \begin{bmatrix}
            \bR^c \\
            \bR^t - \bw^t
        \end{bmatrix},
    \end{equation}
    where $\bbI$ is the identity matrix, the double superscripts denote corresponding tip-channel sub-matrices, and the right-hand side is defined as follows
    \begin{equation}\label{slae_common_rhs}
        \bR^J = \bw_0^J + \Delta t \left( \bbA^{Jc} \bbC^{ct} \bw^t + \bQ^J + \bDeltaSigma^J \right) - \bDeltaL^J, \qquad J = c, t.
    \end{equation}
    The impact of stress layers in the above equation is calculated as $\bDeltaSigma^J = \bbA^{Jc} \bsigma_{h}^{c} + \bbA^{Jt} \bsigma_{h}^{t}$. The vector $\bw_0$ denotes the known fracture width from the previous time step $t - \Delta t$. The fluid leak-off term for the channel elements $\bDeltaL^c$ is calculated using~\eqref{leakoff_discrete_channel}. Calculation of the fluid leak-off $\bDeltaL^t$ is described in~\ref{appendix:tip_leakoff}.
    
    It is worth noting that the elasticity matrix $\bbC$ is dense, while the flux matrix $\bbA$ is sparse. The system~\eqref{slae_without_contact} is the nonlinear system because the matrix $\bbA = \bbA(\bw)$ depends on the current solution. Note that that system of equations~\eqref{slae_without_contact} is suitable only for the case of a non-constrained fracture i.e. $w_{i, j} > {\wmin}_{i, j}$ and $T_n = p$.
    
    If a contact occurs in the $(i, j)$ cell, then the conditions $w_{i, j} = {\wmin}_{i, j}$ and $T_n \neq p$ are satisfied. Hence, the fracture width becomes a known parameter (is equal to the prescribed minimal width ${\wmin}_{i, j}$), while the normal traction ${T_n}_{i, j}$ becomes unknown. Here one can consider an analogy with the tip elements, where the fracture width for the tip elements is determined by the asymptotic solution. Hence, contact elements can be handled similarly, where the fracture width at the constrained elements is defined by the contact condition~\eqref{contact_condition}. Thus, we need to add one more group of elements in addition to the channel and tip variables. This group contains the variables in the elements with active constraint condition~\eqref{contact_condition}. The values corresponding to the contact cells are indicated by the superscript ``d'' (deactivated). Then, given the presence of contact elements, the system of equations~\eqref{slae_without_contact} can be rewritten as follows
    \begin{equation}\label{slae_with_contact}
        \begin{bmatrix*}[r]
            \bbI - \Delta t \bbA^{cc} \bbC^{cc} & -\Delta t \bbA^{ct} & -\Delta t \bbA^{cd} \\
            -\Delta t \bbA^{tc} \bbC^{cc}       & -\Delta t \bbA^{tt} & -\Delta t \bbA^{td} \\
            -\Delta t \bbA^{dc} \bbC^{cc}       & -\Delta t \bbA^{dt} & -\Delta t \bbA^{dd}
        \end{bmatrix*}
        \begin{bmatrix}
            \bw^c \\
            \bp^t \\
            \boldsymbol{T}_n^d
        \end{bmatrix}
        =
        \begin{bmatrix}
            \widetilde{\bR}^c \\
            \widetilde{\bR}^t - \bw^t \\
            \widetilde{\bR}^d - \bwmin^d
        \end{bmatrix},
    \end{equation}
    where the right-hand-side is defined as
    \begin{equation}\label{slae_common_rhs_contact}
        \widetilde{\bR}^J = \bR^J + \Delta t \, \bbA^{Jc} \bbC^{cd} \bwmin^d, \qquad J = c, t, d.
    \end{equation}

\section{Tip leak-off calculation}\label{appendix:tip_leakoff}
    This section aims to describe a simplified approach for calculating the tip leak-off volume based on the midpoint rule. Since a tip element is partially filled, see Figure~\ref{img:tip_element_scheme}, the leak-off term should be integrated only over the internal part of the tip element. According to~\eqref{reynolds_equation} and~\eqref{reynolds_equation_discrete}, the full leak-off is obtained by integration using the midpoint rule
    \begin{equation}
        \mathcal{L}(t) = 2 \Cprime \frac{S}{\Delta x \Delta y} \sqrt{t - t_{0m}},
    \end{equation}
    where $S$ is the area of the $ADEFG$ polygon, $t_{0m}$ is the time at which the fracture front was located at the center of $ADEFG$ polygon. In the case of $\alpha \neq 0$ and $\alpha \neq \pi/2$ the value of $S$ is calculated as $S_{ABC} - S_{DBE} - S_{GFC}$. If $l \leq \Delta x \cos\alpha$ then triangle $DBE$ disappears, while in the case of $l \leq \Delta y \sin\alpha$ the triangle $GFC$ disappears. In the case of $\alpha = 0$ or $\alpha = \pi/2$, the area becomes rectangular. Finally, the value of $S$ is calculated as follows
    \begin{equation}\label{area_partial_filled}
        S = 
        \begin{cases}
            S_\triangle(l) - S_\triangle(l - \Delta x \cos\alpha) - S_\triangle(l - \Delta y \sin\alpha), \quad &\alpha \neq 0, \, \alpha \neq \pi/2 \\
            l \Delta y, \quad &\alpha = 0 \\
            l \Delta x, \quad &\alpha = \pi/2,
        \end{cases}
    \end{equation}
    where $S_\triangle(l) = l^2 / \sin 2 \alpha$ for $l > 0$ and $S_\triangle(l) = 0$ otherwise. The time $t_{0m}$ is implicitly determined by the following equation
    \begin{equation}\label{exposure_time}
        \int_{t_{0m}}^{t} V(\tau) \dint\tau = \frac{l}{2}.
    \end{equation}

\bibliographystyle{elsarticle-num} 
\bibliography{bibliography.bib}
            
\end{document}